\documentclass[%
 reprint,
 amsmath,amssymb,
 aps,
 prd
floatfix,
showkeys
]{revtex4-1}

\usepackage[USenglish]{babel}
\usepackage{slashed}
\usepackage{physics}
\usepackage[normalem]{ulem}
\usepackage{graphicx}% Include figure files
\usepackage{dcolumn}% Align table columns on decimal point
\usepackage{bm}% bold math
\usepackage[usenames]{color}
\usepackage{xcolor}
\usepackage{float}
\usepackage{hyperref}
\usepackage{orcidlink}
\usepackage{verbatim}

\begin{document}

\title{Impact of muons on the bulk viscosity of neutron star matter metamodels}

\author{José Luis Hernández$^{1,2,3}$
\orcidlink{0000-0002-7241-2843}}
\email{hernandez@ice.csic.es}
\author{Cristina Manuel$^{1,2}$
\orcidlink{0000-0002-0024-3366}}
\email{cristina.manuel@csic.es}
\author{Laura Tolos$^{1,2}$
\orcidlink{0000-0003-2304-7496}}
\email{tolos@ice.csic.es} 

\affiliation{$^{1}$Institute of Space Sciences (ICE-CSIC), Campus UAB,  Carrer de Can Magrans, 08193 Barcelona, Spain}
\affiliation{$^{2}$Institut d'Estudis Espacials de Catalunya (IEEC), 08034 Barcelona, Spain}
\affiliation{$^{3}$Facultat de Física, Universitat de Barcelona, Martí i Franquès 1, 08028 Barcelona, Spain.}

\date{\today}

\begin{abstract}
Recent studies invoke a unified description of
different neutron star observables using metamodels,  which parametrize the equation of state of neutron star matter close to nuclear saturation density in terms of few nuclear parameters. In this light, 
the bulk viscosity in the neutrino-transparent regime of dense nuclear matter composed of neutrons, protons and electrons has been recently shown to be mostly sensitive  to the value of the nuclear symmetry energy. As muons are also present  at densities around nuclear saturation, we further analyze in this manuscript their impact on this transport coefficient as a function of the slope $L$ of the symmetry energy. We find that muons introduce both relevant qualitative and quantitative effects in the bulk viscous dissipation.
Increasing  $L$ by a  factor two  has an effect of several orders of magnitude on the (frequency-independent) bulk viscosity.  We also find that for all values of $L$ the frequency-dependent bulk viscosity presents a double peak structure for some values of the density, absent without muons. This also represents changes in orders of magnitude of the viscosity in narrow windows of densities that could be attainable in a neutron star for enough high values of $L$. The double-peak structure is due to the fact that direct Urca processes open up at different densities for electrons and muons, the critical values depending on $L$.
We present a systematic numerical analysis of both second-order transport coefficients, frequency-dependent bulk viscosity, and damping times of density oscillations as a function of the density and the slope, and find when these could be relevant for the dynamics of the merger of neutrons stars.   
\end{abstract}

\maketitle

%%%%%%%%%%%%%%%%%%%%%%%%%%%%%%%%%%%%%%%%%%%%%%%%%%%%%%%%%%%%%%%%%
\section{Introduction}\label{sec1-CM}

Neutron stars (NSs) are one of the most compact objects in the Universe \cite{Glendenning:1997wn}. Structured in different
layers of increasing density, they are generally thought to consist primarily of neutrons, protons, and leptons, although the presence of more exotic degrees of freedom--such as hyperons or even deconfined quark matter--cannot be excluded. The densities realized in these compact objects cannot be reproduced in terrestrial laboratory conditions. Consequently, the behavior of their interiors can only be inferred from theoretical models, validated at far lower densities, requiring extrapolations spanning many orders of magnitude.

The internal structure of NSs is characterized by the equation of state (EoS), which defines their observable properties like masses, radii or cooling \cite{Lattimer:2000nx,MUSES:2023hyz}. The study of the transport properties of nuclear matter at high densities is also of major relevance for the description of NSs. Whereas the EoS fixes the mass–radius relation of
the star, the transport properties—such as viscosities and thermal conductivities—govern its dynamical response to external perturbations \cite{Schmitt:2017efp}. Measurements of the frequencies and damping times of the stellar oscillation modes could, in principle, be used to infer the values
of the various transport coefficients, thereby yielding information on the star’s internal composition, following an approach already employed for other types of stars, such as the
Sun \cite{Christensen-Dalsgaard:2002ney,Lindblom:1983ps,Andersson:1997rn}.

The emergence of gravitational wave (GW) astronomy is transforming the study of NS physics. It has provided
crucial constraints on the EoS after the detection of binary neutron-star mergers ~\cite{Abbott_2018}.
This is achieved by performing numerical simulations of the coalescence of two NSs and
comparing the theoretically computed GW signal with that measured by GW detectors. Until
 now, numerical simulations have neglected transport coefficients \cite{Baiotti:2016qnr,Baiotti:2019sew,Dietrich:2020eud}, with the exceptions of
 \cite{Most:2022yhe,Chabanov:2023blf,Chabanov:2023abq}. One hopes that future implementation of transport effects in these numerical simulations will open a new avenue for constraining transport properties beyond
what is possible through conventional electromagnetic astronomy.

There are strong indications that transport coefficients will play an essential role in modeling NS mergers. In particular, bulk viscosity appears to operate on timescales comparable to those associated with the merger dynamics \cite{Alford:2017rxf}, while the influence of other coefficients depends sensitively on the temperatures attained during the event. This has opened a renewed interest for the study of the bulk viscosity \cite{Sawyer:1989dp,Haensel:1992zz,Haensel:2000vz,Haensel_2001,Alford:2018lhf,Alford_2019,Alford_2021,Alford:2022ufz,Alford_2023,Alford:2023gxq} (see \cite{Harris:2024evy} for a review and a more complete set of references), as it could leave an imprint on the GW emitted in the merger of NSs. 
The GW data  from the binary neutron star GW170817 has recently been used to analyze the dissipative tidal deformability, yielding bounds on the average bulk and shear viscosities of NSs \cite{Ripley:2023lsq}. In fact, both the bulk viscosity associated to nuclear matter with hyperons, and that associated to strange quark matter would result in a phase shift of the GW form
in the inspiral phase of the neutron star merger \cite{Ghosh:2023vrx,Ghosh:2025wfx}. Earlier estimates of the viscosities in nuclear matter suggest that  they do not have an impact in the inspiral phase \cite{Lai_1994,Bildsten1992,Arras2019},  while they can have an effect on the postmerger dynamics. These claims motivate us to have a closer inspection on the viscosities as their values are very much affected by the underlying EoS of nuclear matter. 

The microscopic computation of the bulk viscosity in the  nucleonic case depends both on the electroweak interactions among neutrons, protons and leptons, as well as on
the EoS which dictates the partial densities associated to each particle species. Among the electroweak interactions direct-Urca (dUrca) processes at sufficiently low temperatures are forbidden
below some critical values of the density, the specific value  depending on the particular choice of the EoS
describing the matter, and then the bulk viscosity is dominated by modified-Urca (mUrca) processes, with much lower rates.

In a recent publication \cite{Yang:2025yoo},  see also \cite{Harris:2025ncu},  it has been stressed that the bulk viscosity is highly dependent on the value of the symmetry energy, which quantifies the energy cost of an isospin asymmetry in nuclear matter. The symmetry energy can be parametrized by its value $J$ and its slope $L$ at nuclear saturation density $n_0\approx0.15\, \text{fm}^{-3}$. 
The authors of \cite{Yang:2025yoo} noted that  small changes in the value of $L$ could produce changes in the bulk viscosity of several orders of magnitude. This abrupt change is related to the opening of dUrca processes,  which is highly dependent on the symmetry energy \cite{Cavagnoli:2011ft}.

In this paper we study the bulk viscosity of nuclear matter composed of both neutrons, protons, electrons, and also muons. It had been previously computed in the neutrino-transparent regime in Refs.~\cite{Alford_2023,Alford:2022ufz} and also in the neutrino-trapped regime in~\cite{Alford_2021}. Here we only consider the neutrino-transparent regime, and leave the neutrino-trapped regime for future works.
Muons are allowed by most models of EoSs at densities around saturation density, so that it is relevant to assess their effect. The previous studies were done using two different microscopic models, characterized by different EoS and  different nuclear symmetry parameters. In this work we use a metamodel to describe the EoS of nuclear matter, which allows us to check
the dependence of the viscosity on the slope energy $L$. As in previous studies, we find that small changes in
$L$ lead to big changes in the bulk viscosity, a fact that was already noticed in the models studied in Refs.~\cite{Alford_2023,Alford:2022ufz}. We also find some new big qualitative and quantitative changes. We find that  the critical densities for
the opening of dUrca processes for electrons and muons are different and relatively low for high values of $L$. This has a relevant impact on the value of the frequency-dependent
bulk viscosity that has been unnoticed so far, as we find a double peak resonance  for some values of the densities, which is due to the presence of muons.

This paper is organized as follows. In Sec.~\ref{sec-EOS} we describe the EoS
of the nuclear matter composed of neutrons, protons, electrons and muons, written as an expansion around nuclear saturation density and in terms of nuclear parameters. In Sec.~\ref{sect.nuclearconstraints} we review the constraints that several experiments put on the nuclear parameters. Section~\ref{sec-EWrates} is devoted to review all the electroweak processes that equilibrate the system after an expansion or rarefaction, and give the numerical values of the dUrca and mUrca processes that will be used in our computations. We present the formalism that we use to compute the bulk viscosity in Sec.~\ref{sec-generalbulk}  that  allows us to compute either the second-order transport coefficients that enter in a Burgers equation (Sec.~\ref{section:Burgers_eq}), or the frequency-dependent bulk viscosity (Sec.~\ref{sec-freqbulk}). Our numerical results for the second-order transport coefficients and of the frequency-dependent bulk viscosity are presented in Sec.~\ref{sec.results}, while we evaluate the value of the damping time of density oscillations due to the bulk viscosity in Sec.~\ref{sec-damping}. We end the manuscript with a discussion and conclusion of our results in Sec.~\ref{sec-conclusions}. Details of the computations are given in the Appendices, such as of the thermodynamical relations associated to the metamodel in
Appendix~\ref{sec.thermodynamics}, explicit analytical expressions of the second-order transport coefficients in Appendix~\ref{app.transportcoefficients}, or a discussion of the peak value of the frequency-dependent bulk viscosity when the muon channels are frozen in  Appendix~\ref{sec-app-peak}. We use natural units $k=c=\hbar=1$, except in the figures, where we  plot the bulk viscosity and relaxation times in C.G.S. units. We use
 metric conventions $(+,-,-,-)$.

\section{EoS for nuclear matter}
\label{sec-EOS}

In this section we briefly discuss the metamodeling for the nuclear EoS at zero temperature, which, together with the chemical equilibrium and electric charge neutrality  ($\beta$-equilibrium) constraints, enables us to obtain the thermodynamics needed to compute the transport coefficients. The impact of these metamodels on other neutron star properties, such as the value of masses and radii have been analyzed somewhere else \cite{Margueron:2017eqc,Margueron:2017lup}.

The energy density of nuclear matter including electrons and muons at zero temperature can be written as a sum of the contribution from nucleons and their interactions based on a Taylor expansion around the nuclear saturation density and symmetric matter (when the proton fraction $X_p\equiv n_p/n_B$ equals $1/2$) and the one from leptons described as ideal Fermi gases. Defining the variables
\begin{equation}
\label{exp-parameter}
x\equiv\frac{n_B-n_0}{3n_0},  \qquad \beta\equiv 1-2X_p  ,
\end{equation}
and expanding around zero for these variables, we have

\begin{equation}
\label{eq.energy_density}
\begin{split}
\varepsilon(n_B, X_p, n_e, n_\mu)
&= n_B
\left( m - B_{\rm sat} + \frac{K}{2}x^2 + \frac{Q}{6}x^3 \right) \\
&+ n_B S(n_B)\,\beta^2
+ \frac{\mu_e^4}{4\pi^2} \\
&
+ \frac{1}{8\pi^2}
\Bigg\{ \mu_\mu \sqrt{\mu_\mu^2 - m_\mu^2} 
\left( 2\mu_\mu^2 - m_\mu^2 \right) \\
&- m_\mu^4
\ln\left[ \left(\mu_\mu + \sqrt{\mu_\mu^2 - m_\mu^2}\right)/m_\mu
\right]
\Bigg\},
\end{split}
\end{equation}
where $\mu_e$ and $\mu_\mu$ are the electron and muon chemical potential, respectively, and 
$m_\mu\approx105$ MeV is the muon mass. We neglect the electron mass, as it is much smaller than $\mu_e$. 

The first term is the energy per baryon of symmetric nuclear matter where $m$ is the nucleon mass in vacuum, $B_{\rm sat}$ is the binding energy, $K$ and $Q$ are  the incompressibility and the skewness in symmetric nuclear matter, respectively.
The second expression considers the asymmetric part of the expansion with $S(n_B)$ defined as
\begin{eqnarray}
    S(n_B)&\equiv& J+Lx +\frac{K_{\rm sym}}{2}x^2+\frac{Q_{\rm sym}}{6}x^3,
\end{eqnarray}
with $J$ the symmetry energy at saturation density, $L$ the slope, $K_{\rm sym}$ the incompressibility and $Q_{\rm sym}$ the skewness in asymmetric matter. We elaborate on their values according to nuclear experiments constraints at nuclear saturation in the next subsection. The third and fourth terms are the electron and muon contributions to the energy density, respectively~\cite{Wen_2005}. Note that we can use the following free thermodynamic relations for the electron density 
\begin{equation}
    n_e=\frac{\mu_e^3}{3\pi^2},
\end{equation}
and for the muon density
\begin{equation}
    n_\mu=\frac{1}{3\pi^2}(\mu_\mu^2 -m_\mu^2)^{3/2} \ ,
\end{equation}
in Eq.~\eqref{eq.energy_density} to write the chemical potentials of leptons in terms of their number densities.

Finally, we consider the chemical equilibrium and electric charge neutrality conditions in Eq.\eqref{eq.energy_density} to determine the chemical potentials and number densities for our calculations. Using the following expressions of the neutron and proton number densities, $n_n$ and $n_p$, respectively,
\begin{equation}
    \begin{aligned}
        n_p&=n_Q+n_e+n_\mu, \\
        n_n&=n_B-n_Q-n_e-n_\mu, 
    \end{aligned}
\end{equation}
with the baryon and electric charge number densities defined as follows
\begin{equation}
    \begin{aligned}
        n_B&\equiv n_n+n_p, \\
        n_Q&\equiv n_p -n_e-n_\mu, 
    \end{aligned}
\end{equation}
we fix $n_B$ and set $n_Q=0$. In this way, we have two unknown variables, namely, $n_e$ and $n_\mu$, which together with the  $\beta$-equilibrium conditions  $\mu_e=\mu_n -\mu_p$ and  $\mu_\mu=\mu_e$, enable us to determine the remaining variables. Thus, given the baryon number density we are able to determine the thermodynamics of the system in $\beta$-equilibrium. 

\begin{figure}[b]
\includegraphics[width=\linewidth]{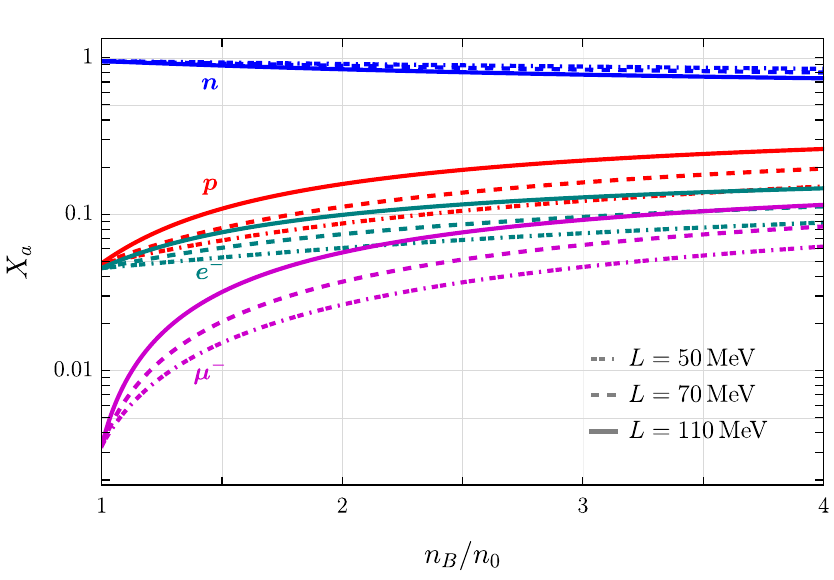}
\caption{Particle fractions $X_a$ as a function of the baryon number density $n_B$ divided by nuclear saturation density $n_0$ at $L=50$ MeV (dotted-dashed lines), $L=70$ MeV (dashed lines) and $L=110$ MeV (continuous lines). The different curves are obtained at $J=32$ MeV, $K=240$ MeV and $K_{\rm sym}=Q=Q_{\rm sym}=0$.}
\label{fig:Xa_with_density_J32_K240_Ksym0_varying_L_and_temperature}
\end{figure}

In Fig.~\ref{fig:Xa_with_density_J32_K240_Ksym0_varying_L_and_temperature} we show the particle fraction $X_a \equiv n_a/n_B$ as a function of the baryon number density divided by the nuclear saturation density in $\beta$-equilibrium with $a=n,\,p,\,e,\,\mu$, hereafter we consider $n_0=0.15\, \text{fm}^{-3}$\cite{PhysRevC.102.044321}. Our calculations are performed for densities from $n_B=n_0$ up to $4n_0$, and  at $L=50$, $L=70$ MeV and $L=110$ MeV, while fixing $J=32$ MeV, $K=240$ MeV and $K_{\rm sym}=Q=Q_{\rm sym}=0$ MeV. The values for the nuclear parameters are discussed in the next subsection. Particularly, for the cases considered in this plot, we observe muons appear slightly below nuclear saturation density. Proton, electron and muon fractions  increase with density, more strongly for larger values of $L$,  while the neutron fraction decreases accordingly.

\subsection{Constraints on nuclear parameters from experiments} \label{sect.nuclearconstraints}

Several constraints on the values of the previously defined nuclear parameters can be extracted from nuclear experiments (see \cite{particles6010003} for a comprehensive review).

As for the parameters associated to symmetric nuclear matter, the so-called isoscalar parameters $B_{\rm sat}$, $K$ and $Q$, only $B_{\rm sat}$ is well constrained. Measurements of density distributions \cite{DeVries:1987atn} and nuclear masses \cite{AUDI2003337} allowed for the determination of $B_{\rm sat}=-16 \pm 1$ MeV at the nuclear saturation density $n_0=0.15-0.16 \ {\rm fm}^{-3}$.  As for the incompressibility at saturation density $K$, the extraction comes from the analysis of isoscalar giant monopole resonances
in heavy nuclei. However, the determination of its value is complicated and ambiguous, with possible values in the range of $K\sim$ 200-300 MeV (see for example \cite{Blaizot:1980tw,Piekarewicz:2003br,Khan:2012ps}).  Moreover, $Q$ is not well constrained, with a band of uncertainty of a few hundreds of MeVs.

With regard to the parameters $J$, $L$, $K_{\rm sym}$ and $Q_{\rm sym}$, also called isovector parameters as they are associated to the asymmetric part of the EoS, these could be determined from experiments involving isospin diffusion measurements \cite{Chen:2004si}, isobaric analog states \cite{Danielewicz:2008cm}, isoscaling \cite{Shetty:2007zg}, analysis of giant  \cite{Garg:2006vc} and pygmy resonances \cite{Klimkiewicz:2007zz,Carbone:2010az},  production of pions \cite{Li:2004cq} and kaons \cite{Fuchs:2005zg,Hartnack:2011cn,Song:2020clw} in heavy-ion collisions or data on neutron skin thickness of heavy nuclei \cite{Brown:2000pd,Horowitz:2000xj,Horowitz:2001ya,Centelles:2008vu,PREX:2021umo,CREX:2022kgg}. Whereas $J$ is relatively well determined with values around $\sim 30$ MeV for nuclear saturation density, $L$ and $K_{\rm sym}$ are still poorly known. In particular, $L$ could vary from few tenths of MeV up to $\sim$120 MeV, this last value inferred from the parity-violating electron scattering neutron skin
experiment in $^{208}$Pb \cite{PREX:2021umo}. Moreover, $K_{\rm sym}$ or $Q_{\rm sym}$ are barely constrained. For the discussion on the values for the isovector parameters and possible correlations among them,  we refer the reader to  Ref.~\cite{particles6010003}. In most part of this article  we will neglect the effect of $K_{\rm sym}$, $Q$, and $Q_{\rm sym}$ as there is not much we know of these parameters and set them to zero unless otherwise stated.

\section{Electroweak rates relevant for the bulk viscosity}
\label{sec-EWrates}

Neutron stars are driven out of chemical equilibrium in a  volume expansion and/or compression, thereby deviating values of chemical potentials and number densities of the medium from their equilibrium values. The electroweak interactions provide a mechanism for the system to reestablish chemical equilibrium in the medium at the dynamical timescales of the system. As a starting point, the evolution equations of the number densities of neutrons, protons, electrons and muons out of equilibrium can be expressed as 
\begin{equation}
\label{eq-evolutiondensities}
    \begin{split}
        \nabla_\mu(u^\mu n_n(t))&= \Gamma_{\text{LC},e}-\Gamma_{\text{ND},e} +\Gamma_{\text{LC},\mu}-\Gamma_{\text{ND},\mu}, \\
\nabla_\mu(u^\mu n_p(t))&= \Gamma_{\text{ND},e}- \Gamma_{\text{LC},e}+\Gamma_{\text{ND},\mu}- \Gamma_{\text{LC},\mu}, \\
\nabla_\mu(u^\mu n_e(t))&= \Gamma_{\text{ND},e}- \Gamma_{\text{LC},e}
+\Gamma_{\text{MD}}, \\
\nabla_\mu(u^\mu n_\mu(t))&= \Gamma_{\text{ND},\mu}- \Gamma_{\text{LC},\mu}-\Gamma_{\text{MD}}, 
    \end{split}
\end{equation}
respectively, 
where the electroweak transition rates $\Gamma$ act as source or sink terms to restore equilibrium number densities. Here we use the subscripts $\text{LC}$ for lepton capture, $\text{ND}$ for neutron decay, $\text{MD}$ for the muon decay and $l=e,\mu$ denotes the lepton involved in the process. The lepton capture and neutron decay electroweak processes receive contributions either from dUrca processes or mUrca processes. The dUrca processes correspond to the reactions 
\begin{equation}
    \begin{aligned}
        &n\to p+l+\bar{\nu}_l, \\
        &p+l \to n+\nu_l,
    \end{aligned}
\end{equation}
and contribute to the neutron decay and the lepton capture, respectively. Here $\nu_l$ ($\bar{\nu}_l$) denotes neutrinos (antineutrinos) associated to the lepton $l$. 

When dUrca processes are forbidden by momentum conservation, the bulk viscosity is then generated by the mUrca processes
\begin{eqnarray}
    && n+ N\to p+N+l+\bar{\nu}_l, \\
    && p+ N +l \to n + N +\nu_{l},
\end{eqnarray}
where $N=n,\,p$ is a spectator nucleon that guarantees momentum conservation. The different choices of the spectator nucleon are known as the neutron or proton branches of the mUrca processes~\cite{Haensel_2001}. 

\begin{figure}[t]
\includegraphics[width=\linewidth]{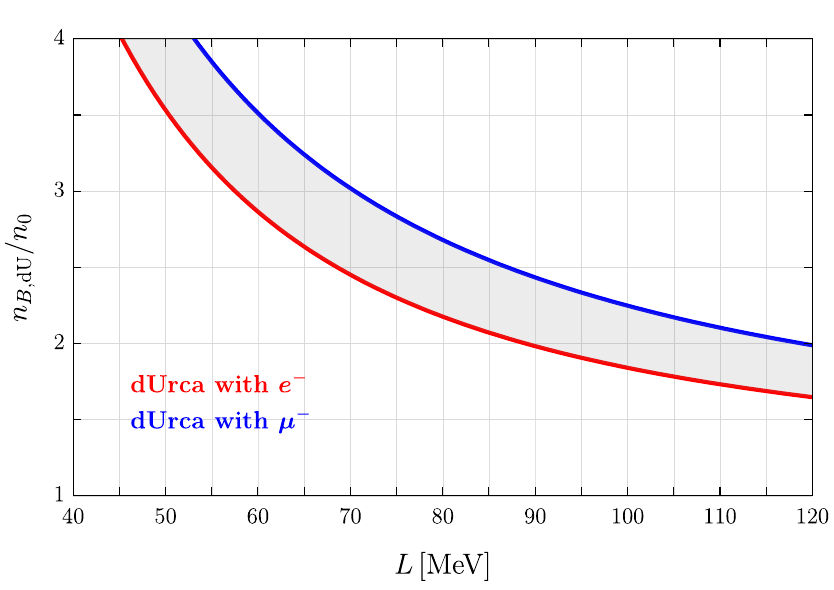}
\caption{Density threshold of the dUrca processes $n_{B,\text{dU}}$ in
multiples of the saturation density $n_0$ as a function of the symmetry slope. Only for densities above the red/blue curve dUrca processes
for electrons/muons are kinematically allowed. The different curves are
obtained at $J = 32$ MeV
and $K = 240$ MeV, and $K_{\rm sym} = Q = Q_{\rm sym} = 0$.}
\label{fig:thrdU_with_slope_nuclear_matter_J32_K240_Ksym0}
\end{figure}

The critical density where the dUrca processes are allowed depends on the symmetry energy, as this controls the proton and electron densities, and thus, it changes according to the EoS.  In 
Fig.~\ref{fig:thrdU_with_slope_nuclear_matter_J32_K240_Ksym0}, we display the density threshold $n_{B,\text{dU}}$ of dUrca processes as a function of the slope $L$ from $40$ MeV up to high values around $120$ MeV predicted by the PREXII experiment~\cite{PREX:2021umo}.
%\cite{Reed_2021}.
As can be seen, dUrca processes with electrons have a lower threshold density as compared to the ones with muons. Our results are in qualitative and quantitative agreement with previous studies~\cite{Cavagnoli_2011, Provid_ncia_2019, Malik_2022}, reproducing the decreasing trend of the density threshold with increasing $L$.  As $L$ increases, the neutron-proton asymmetry above saturation density diminishes, and, hence, the threshold density for the dUrca process shifts to lower values. 

\begin{figure}[t]
\includegraphics[width=\linewidth]{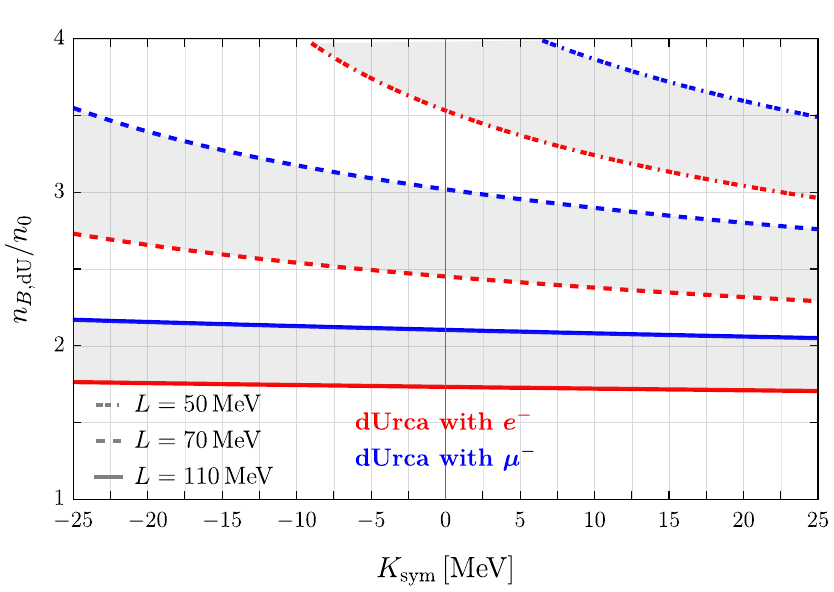}
\caption{Density threshold of the dUrca processes $n_{B,\text{dU}}$ in
multiples of the saturation density $n_0$ as a function of the incompressibility $K_{\rm sym}$. Only for densities above the red/blue curve dUrca processes
for electrons/muons are kinematically allowed. We vary $L$ as follows: $L=50$ MeV (dashed-dotted lines), $L=70$ MeV (dashed lines) and $L=110$ MeV (continuous lines). The different curves are
obtained at $J = 32$ MeV
and $K = 240$ MeV, and $ Q = Q_{\rm sym} = 0$.}
\label{fig.thrdU_with_curvature_nuclear_matter_J32_K240}
\end{figure}

In Fig.~\ref{fig.thrdU_with_curvature_nuclear_matter_J32_K240} we show the critical density for the opening of dUrca process as a function of the (poorly) constrained $K_{\rm sym}$ parameter at $L=50$ MeV, $L=70$ MeV, and $L=110$ MeV. 
Only positive values of $K_{\rm sym}$ produce lower thresholds of both dUrca processes including electrons and muons compared to the scenario where $K_{\rm sym}=0$. In addition, these thresholds slightly decrease as $K_{\rm sym}$ increases its value, the decreasing trend is stronger as we consider small values of $L$. When $K_{\rm sym}<0$ we notice the opposite effect.

Hereafter, we consider  small deviations of chemical equilibrium in the system. To this end, we define the following chemical imbalances
\begin{equation}
\begin{aligned}
\mu_1 &\equiv \delta\mu_n - \delta\mu_p - \delta\mu_e \ , \\
\mu_2 &\equiv \delta\mu_n - \delta\mu_p - \delta\mu_\mu \ ,
\end{aligned}
\label{eq.chemicalimbalances}
\end{equation}
where $\delta \mu_a$ with $a=n,p,e,\mu$ are out-of-equilibrium deviations of the chemical potentials small enough to describe subthermal bulk-viscous effects so that $\mu_1,\mu_2\ll 2\pi T$. The chemical imbalances  represent two independent degrees of freedom to describe the system out of equilibrium. In this way, the following differences of electroweak rates can be computed at linear order in $\mu_1$ and $\mu_2$
\begin{equation}
\begin{aligned}
    &\Gamma_{\text{ND},e}- \Gamma_{\text{LC},e} \approx \lambda_1 \mu_1, \\
    &\Gamma_{\text{ND},\mu}- \Gamma_{\text{LC},\mu}\approx \lambda_2\mu_2, \\
    &\Gamma_{\rm MD}\approx \lambda_3\mu_3,
\end{aligned}
    \label{eq.rates_differences}
\end{equation}
with $\mu_3= \mu_1-\mu_2$. The electroweak coefficients $\lambda_1$ and $\lambda_2$ can be expressed as follows
\begin{equation}
    \begin{aligned}
    \lambda_1 &= \lambda_e +\lambda_{ne}+\lambda_{pe}, \\
    \lambda_2 &= \lambda_\mu+\lambda_{n\mu}+\lambda_{p\mu} , 
    \end{aligned}
\end{equation}
where $\lambda_l$ and $\lambda_{Nl}$ are the contributions from the dUrca and the mUrca processes, respectively, where subscripts denote the branch of the spectator nucleon $N$ and $l$ the lepton involved in the process. 

The coefficients for dUrca processes $\lambda_l$ in the low-temperature regime and in the Fermi surface approximation, this is, considering that only the particles near the Fermi surface take part in the electroweak processes are given by~\cite{Alford_2023}
\begin{eqnarray}
    \lambda_{l} &\approx& \frac{17}{480 \pi} G^2 T^4 \mu_n^* \Theta\left(p_{F,l}+p_{F,p}-p_{F,n} \right) \nonumber \\ &\times&\left(p_{F,p}^2 +p_{F,l}^2 +2\mu_l\mu_p^*-p_{F,n}^2 \right), \label{eq.dUEW_rate}
\end{eqnarray}    
with $G\equiv G_{F}\cos{\Theta_C}(1+g_A)$, $G_F=1.166\times10^{-5}\,\text{GeV}^{-2}$ the Fermi coupling constant, $\Theta_C\approx 13.02^\circ$ is the Cabbibo angle, $g_A=1.26$ is the axial-vector coupling constant, the effective nucleon chemical potential 
$\mu_{N}^*=\left(k_{F,N}^2+(m_{N}^*)^2\right)^{1/2}$,
the effective nucleon mass $m_N^*\approx (0.7\pm0.1)m_N$ with $m_N\approx 940$
MeV at nuclear saturation density, $\Theta$ denotes the Heaviside function and $p_F$ is the Fermi momentum of the particles. 

For the mUrca processes, we employ the reaction rates coming from a nucleon-nucleon interaction that consists of a long-range one-pion-exchange tensor contribution and a short-range piece parametrized with nuclear Fermi liquid (also known Landau) parameters~\cite{Friman1979}. Within this framework, we use the expressions for $\lambda_{Nl}$ shown in Ref.~\cite{Alford_2019} (see also Ref.~\cite{Haensel_2001} for simplified versions of these expressions) 
\begin{equation}
    \begin{split}
    \lambda_{ne}&\approx\frac{367}{1152\pi^3}G_{F}^2 \cos^2{\Theta_C}g_A^2f_{\pi NN}^4  \\
    &\times\frac{m_n^{*3}m_p^*}{m_\pi^4}\frac{p_{Fn}^4p_{Fp}}{(p_{Fn}^2+m_\pi^2)^2} \vartheta_{ne} T^6, 
    \end{split}
\end{equation}
\begin{equation}
    \begin{split}
    \lambda_{n\mu}&\approx\frac{367}{1152\pi^3}G_{F}^2 \cos^2{\Theta_C}g_A^2f_{\pi NN}^4  \\
    &\times\frac{m_n^{*3}m_p^*}{m_\pi^4}\frac{p_{F\mu}p_{Fn}^4p_{Fp}}{p_{Fe}(p_{Fn}^2+m_\pi^2)^2} \vartheta_{n\mu} T^6,
    \end{split}
\end{equation}
\begin{equation}
    \begin{split}
    \lambda_{pe}&\approx\frac{367}{8064\pi^3}G_{F}^2 \cos^2{\Theta_C}g_A^2f_{\pi NN}^4 \\
    &\times \frac{m_n^*m_p^{*3}}{m_\pi^4}\frac{p_{Fn}(p_{Fn}-p_{Fp})^4}{[(p_{Fn}-p_{Fp})^2+m_\pi^2]^2} \vartheta_{pe}T^6,
    \end{split}
\end{equation}
\begin{equation}
\begin{split}
\lambda_{p\mu}&\approx\frac{367}{8064\pi^3}G_{F}^2 \cos^2{\Theta_C}g_A^2f_{\pi NN}^4  \\
&\times\frac{m_n^*m_p^{*3}}{m_\pi^4}\frac{p_{F\mu}p_{Fn}(p_{Fn}-p_{Fp})^4}{p_{Fe}[(p_{Fn}-p_{Fp})^2+m_\pi^2]^2} \vartheta_{p\mu}T^6,
\end{split}
\end{equation}    
where the functions $\vartheta_{nl}$ and $\vartheta_{pl}$ are given by
\begin{equation}
    \vartheta_{nl}\equiv
    \begin{cases}
        1 & \text{if } p_{Fn}>p_{Fp}+p_{Fl},\\
        1-\frac{3}{8}\frac{(p_{Fp}+p_{Fl}-p_{Fn})^2}{p_{Fp}p_{Fl}} & \text{if } p_{Fn}<p_{Fp}+p_{Fl},
    \end{cases}
\end{equation}

\begin{equation}
\vartheta_{pl} \equiv
\begin{cases}
0,
& \text{if } p_{Fn} > 3p_{Fp} + p_{Fl}, \\[1ex]

\dfrac{(3p_{Fp} + p_{Fl} - p_{Fn})^2}{p_{Fn} p_{Fl}},
& \text{if }
\begin{aligned}
& p_{Fn} > 3p_{Fp} - p_{Fl}, \\
& p_{Fn} < 3p_{Fp} + p_{Fl},
\end{aligned}
\\[3ex]

4\,\dfrac{3p_{Fp} - p_{Fn}}{p_{Fn}},
& \text{if }
\begin{aligned}
& 3p_{Fp} - p_{Fl} > p_{Fn}, \\
& p_{Fn} > p_{Fp} + p_{Fl},
\end{aligned}
\\[3ex]

2 + \dfrac{3(2p_{Fp} - p_{Fn})}{p_{Fl}} & \text{if } 
p_{Fn} < p_{Fp} + p_{Fl}.
\\
\hspace{0.275cm}- \dfrac{3(p_{Fp} - p_{Fl})^2}{p_{Fn} p_{Fl}},

\end{cases}
\end{equation}    
where $f_{\pi NN} \approx1$ is the $\pi N$-interaction constant in the $p$-state in the one-pion-exchange model potential~\cite{Haensel_2001}, $m_\pi\approx 139.57$ MeV is the mass of the charged pions $\pi^{\pm}$ and the Fermi momentum of the particles holds $p_{F,a}=(3\pi^2 n_a)^{1/3}$. 

Lastly, for the leptonic channel, that is the muon decay, corresponding to the process
\begin{equation}
\mu \rightarrow e + {\bar \nu}_e + \nu_{\mu}   \ . 
\end{equation}
The coefficient $\lambda_3$ can be computed as follows
\begin{equation}
    \lambda_3 = \frac{\partial \Gamma_{\rm MD}}{\partial \mu_3}\bigg|_{\mu_3=0},
\end{equation}
where the full expression of $\Gamma_{\rm MD}$ is given in Eq.(17) of Ref.~\cite{Alford_2023}.
However, this process is subdominant with respect to $\lambda_1$ and $\lambda_2$~\cite{Alford_2023}. According to Ref.~\cite{Harris:2024evy} (see Fig. 8), $\lambda_3/\lambda_1$ and $\lambda_3/\lambda_2$ are approximately $0.5\times10^{-7}$ at $n_B=n_0$ and $T=1$ MeV. Hereafter, we will neglect it in our numerical results.
\begin{figure}[t]
\includegraphics[width=0.483\textwidth]{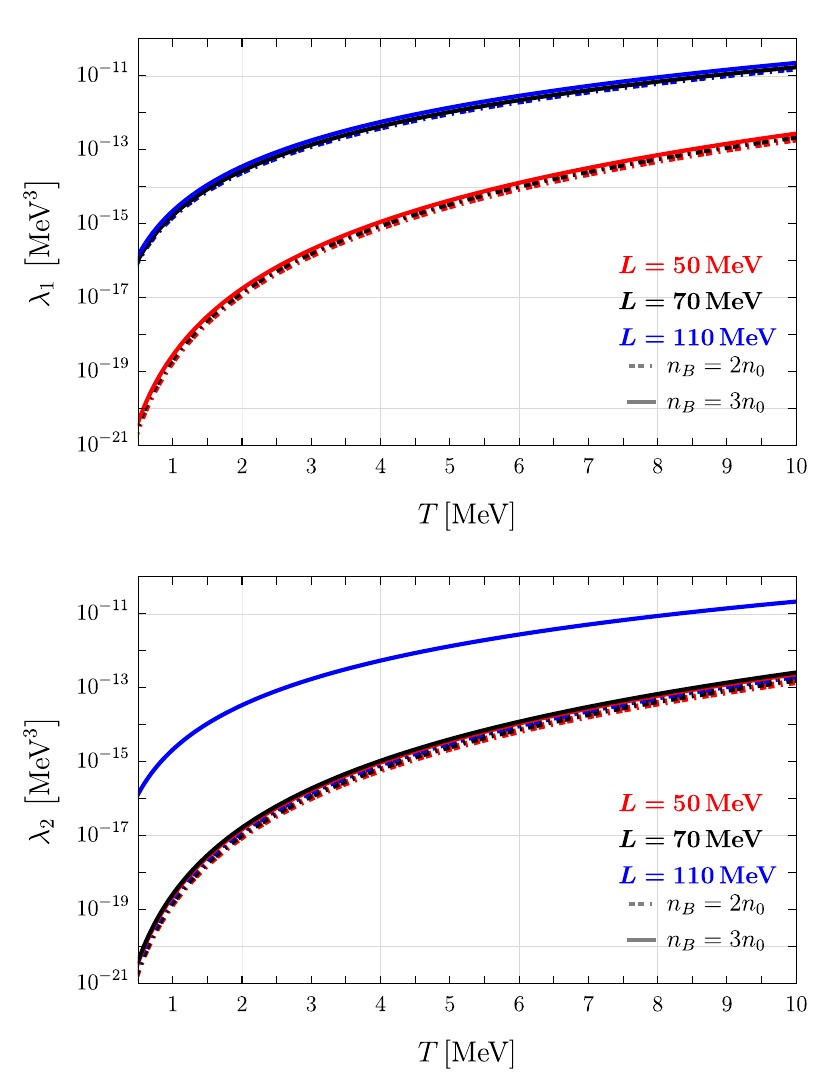}
\caption{Electroweak coefficients $\lambda_1$ (upper panel) and $\lambda_2$ (bottom panel) as a function of temperature $T$ in log scale at $n_B=2 n_0$ (dashed-dotted lines) and $n_B=3n_0$ (solid lines). The red, black and blue curves represent $L=50$ MeV, $L=70$ MeV and $L=110$ MeV, respectively. The different curves are obtained at $J=32$ MeV, $K=240$ MeV and $K_{\rm sym}=Q=Q_{\rm sym}=0$.}
\label{fig:temperature_lambdas_with_J32_K240_varying_L_and_density}
\end{figure}

In Fig.\ref{fig:temperature_lambdas_with_J32_K240_varying_L_and_density} we display the electroweak rates $\lambda_1$ (upper panel) and $\lambda_2$ (bottom panel) as a function of temperature in log-scale and at $n_B=2n_0$ (dashed-dotted lines) and $n_B=3n_0$ (solid lines), for values of the slope  $L=50$ MeV (red curves), $L=70$ MeV (black curves) and $L=110$ MeV (blue curves). We see little dependence on $n_B$ or $L$ (several curves are almost on top of each other) for lower values of $L$, except when either the value of $L$ or of $n_B$ is high enough so as to open the dUrca processes. Then the rates change by several orders of magnitude. 

\begin{figure}[b]
\includegraphics[width=0.483\textwidth]{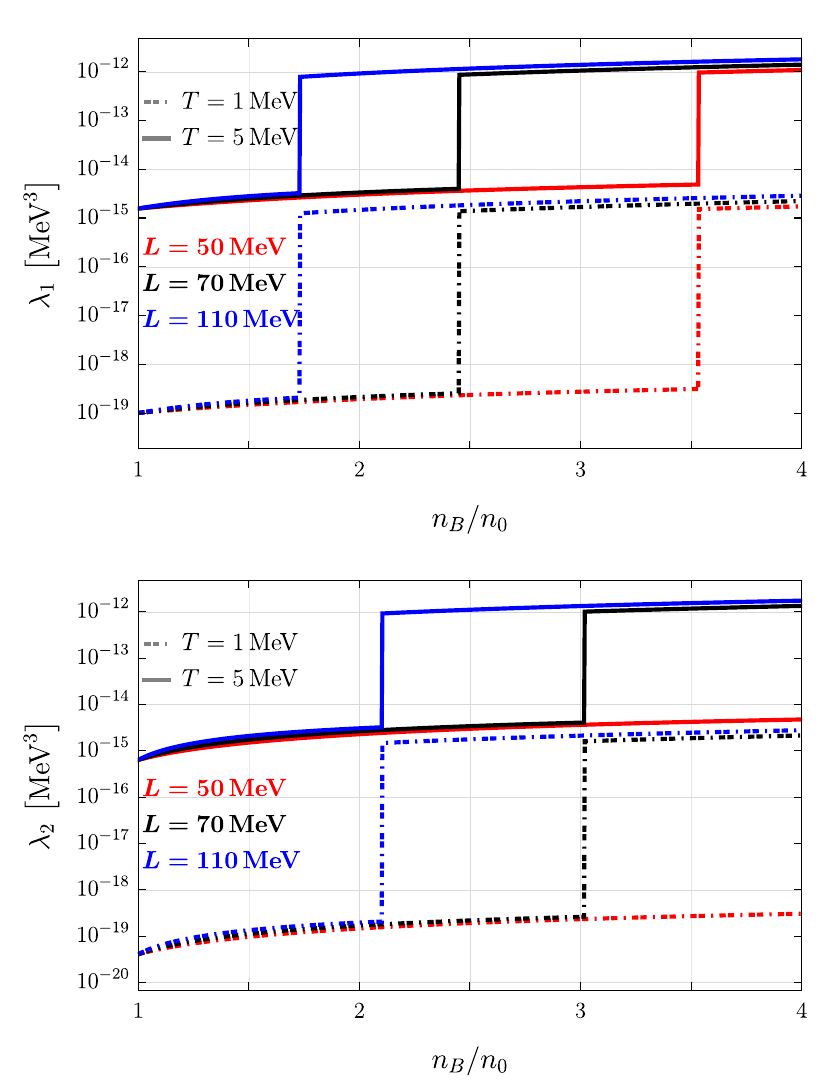}
\caption{Electroweak rates $\lambda_1$ (upper panel) and $\lambda_2$ (lower panel) as a function of baryon number density $n_B$ (in multiples of the saturation density) in log scale at $T=1$ MeV (dashed-dotted lines) and $T=5$ MeV (continuous lines). The red, black and blue curves represent $L=50$ MeV, $L=70$ MeV and $L=110$ MeV, respectively. The different curves are obtained at $J=32$ MeV, $K=240$ MeV and $K_{\rm sym}=Q=Q_{sym}=0$.}
\label{fig:density_lambdas_with_J32_K240_varying_L_and_temperature}
\end{figure}
In Fig.~\ref{fig:density_lambdas_with_J32_K240_varying_L_and_temperature} we display the density profiles of the electroweak rates: $\lambda_1$ (upper panel) and $\lambda_2$ (lower panel). In this case, we evaluate electroweak coefficients at $T=1$ MeV (dashed-dotted curves) and $T=5$ MeV (continuous curves) and also vary $L$ as follows: $L=50$ MeV (red lines), $L=70$ MeV (black lines), and $L=110$ MeV (blue lines). First, electroweak rates at $T=5$ MeV are orders of magnitude higher than at $T=1$ MeV because they receive contributions proportional to $\propto T^4$ from dUrca processes and $\propto T^6$ from mUrca processes. In addition, at a fixed temperature we observe an abrupt increase that occurs at the density threshold of dUrca processes.    For a fixed $L$ this jump occurs  at lower densities for dUrca processes with electrons ($\lambda_1$, upper panel) than dUrca processes with muons ($\lambda_2$, lower panel),  whereas the density threshold for each dUrca is reduced as the value of $L$ increases, as also observed in 
Fig.~\ref{fig:thrdU_with_slope_nuclear_matter_J32_K240_Ksym0}.

It is worth to mention we are using the electroweak rates of dUrca and mUrca processes using the Fermi surface approximation.
It is well-known   
more precise calculations of these rates without the Fermi surface approximation do not produce abrupt steps on the rates as those considered here {~\cite{Alford_2019, Alford:2024xfb}.
We leave these improvements on the rates and its possible impact on the bulk viscosity for future research.

Let us mention that the vast majority of computations of the bulk viscosity in neutron stars use 
the FS approximation, as this allows for analytical computations of the different rates. According
to the results of Ref.~\cite{Alford_2019} (see Fig. 2), in the regime of temperature between 1 to 10 MeV, the FS is still a reasonable approximation, which however differs from the exact treatment of the electroweak rates. As soon as the temperature goes up, the Fermi sea is blurred.
One expects then  the direct Urca density threshold to also blur, exactly as it happens to the maximum value of bulk viscosity as a function of the temperature, as found in Ref.~\cite{Alford_2019}.
This is also the effect we may expect on our results. We only expect refinements on the numerical values we provide, but the main result we discuss would still be there.

\section{Bulk viscosity in nuclear matter including electrons and muons}
\label{sec-generalbulk}
The bulk viscosity in nuclear matter has been addressed in several studies considering different phases at ultrahigh density, either neutrino-transparent and neutrino-trapped matter,  both considering the subthermal or suprathermal regimes. For a review of these studies and references see Ref.~\cite{Harris:2024evy}. At densities close and below to the nuclear saturation density, neutron stars are expected to be composed by neutrons, protons and electrons.
For astrophysical applications, it is customary to give an effective frequency-dependent bulk viscosity. These effective descriptions can also be formulated within
second-order hydrodynamics,  with the help of an Israel-Stewart equation
~\cite{Yang:2023ogo, Yang:2023ozd}.

In this work, we analyze the bulk viscosity of neutrino-transparent nuclear matter including electrons and also muons. The latter can also be generated in the system around and a few times the nuclear saturation density for realistic EoSs and have an important effect on the chemical equilibration of the system. The frequency-dependent bulk viscosity in this medium has been studied using the constitutive relation at first-order hydrodynamics in the transparent-neutrino regime~\cite{Alford_2023,Alford:2022ufz} and also at high enough temperatures for neutrinos to get trapped in the system~\cite{Alford_2021}. Later on, it was realized that the description of these
effective frequency-dependent bulk viscosity could be attained in second-order hydrodynamics, and that the bulk-viscous pressure in neutrino-transparent nuclear matter with electrons and muons follows a Burgers equation~\cite{Gavassino_2023}.

In this section we review these two formalisms, that we will finally use for
the numerical evaluation of the different transport coefficients. 

\subsection{Burgers equation fulfilled by the bulk viscous pressure}
\label{section:Burgers_eq}

In this subsection we derive the dynamical  evolution  of the bulk-viscous pressure in nuclear matter, which corresponds to a Burgers equation when there are both electrons and muons, as first noted in \cite{Gavassino_2023}, see also \cite{Hernandez:2025zxw}. We also
determine the associated second-order transport coefficients.

Close to chemical equilibrium, one can linearize Eqs.\eqref{eq-evolutiondensities} to reach to
\begin{equation}
\begin{aligned}
    \nabla_\mu(u^\mu n_n(t))&\approx -\lambda_1 \mu_1(t) -\lambda_2\mu_2(t), \\
    \nabla_\mu(u^\mu n_p(t))&\approx \lambda_1 \mu_1(t) +\lambda_2\mu_2(t), \\
    \nabla_\mu(u^\mu n_e(t))&\approx \lambda_1\mu_1+\lambda_3[\mu_1(t) -\mu_2(t)], \\
     \nabla_\mu(u^\mu n_\mu(t))&\approx\lambda_2 \mu_2(t) -\lambda_3[\mu_1(t)-\mu_2(t)]. 
\end{aligned}
\end{equation}
The number densities are modified with respect to their values in chemical equilibrium by linear deviations of the chemical potentials
\begin{equation}\label{eq.numberdensity}
    n_a=n_a^{(\rm eq)} +\sum_{b} \chi_{a}^b \delta \mu_b,
\end{equation}
where $n_a^{(\rm eq)}$ is the equilibrium density of the particle species $a$
with the matrix of isothermal susceptibilities defined as  
\begin{equation}\label{eq.isosusceptibility}
    \chi_{a}^b=\frac{\partial n_a}{\partial\mu_b}\bigg|_{\mu_c \neq \mu_b,T,\delta\mu_b=0},
\end{equation}
with $a,\,b,\,c,\,=n,\,p,\,e,\,\mu$. The susceptibilities can also be computed at constant entropy per baryon (adiabatic susceptibilities) valid to 
%{\color{violet} (Laura: "when neglected" instead of "to neglect"?)}
neglect the heat flow between adjacent fluid elements~\cite{Alford_2019}. For a zero-temperature EoS, as in this work, both susceptibilities match.

Using Eq.~\eqref{eq.numberdensity} and the evolution equations of the different particle densities,  we get an expression to study the evolution in time of $\mu_\alpha$, with $\alpha =1,2$,
\begin{equation}\label{eq.evolmualpha}
    u^\mu \nabla_\mu  \mu_\beta= M_\beta^{\,\,a} \left(\lambda_a^{\,\,\alpha} \mu_\alpha -n_a ^{(\rm eq)}\theta \right),
\end{equation}
where $\theta \equiv \nabla_\mu u^\mu$ is the expansion rate. We collect the inverse  susceptibilities in the two vectors
\begin{equation} \label{eq.matrixM}
\begin{split}
    M_1^{\,\,a}\equiv \left(\chi^{-1}\right)^{\,\,a}_n-\left(\chi^{-1}\right)^{\,\,a}_p-\left(\chi^{-1}\right)^{\,\,a}_e, \\
    M_2^{\,\,a}\equiv \left(\chi^{-1}\right)^{\,\,a}_n-\left(\chi^{-1}\right)^{\,\,a}_p-\left(\chi^{-1}\right)^{\,\,a}_\mu, 
\end{split}
\end{equation}
as well as the different electroweak rates in the matrix 
\begin{align}
\lambda_a^{\,\,\alpha}&\equiv\begin{pmatrix}-\lambda_{1} & -\lambda_{2}\\
\lambda_{1} & \lambda_{2}\\
\lambda_{1}+\lambda_{3} & -\lambda_{3}\\
-\lambda_{3} & \lambda_{2}+\lambda_{3}
\end{pmatrix},
\end{align}
with the indexes $\alpha,\beta=1,2$ and $a=n,p,e,\mu$ (in this order).
Thus, Eq.~\eqref{eq.evolmualpha} can be written as an evolution equation of the chemical imbalances 
\begin{equation}\label{eq.master_chemical}
    \tau^{\,\, \alpha}_\beta u^\mu\nabla_\mu \mu_\alpha+\mu_\beta = -\theta b_\beta ,
\end{equation}
with 
\begin{equation}\label{eq.tau}
  \left(\tau^{-1}\right)_\beta^{\,\,\alpha}\equiv-M_\beta^{\,\, a} \lambda_a^{\,\, \alpha},
\end{equation}
and
\begin{equation}\label{eq.bbeta}
    b_\beta=\tau_\beta^{\,\,\alpha} M^{\,\,a}_\alpha n_a^{(\rm eq)} .
\end{equation}

Using Eq.~\eqref{eq.master_chemical} and given that the bulk scalar $\Pi$ defined as $\Pi\equiv P-P_0$ (where $P$ is the total pressure and $P_0$ is the pressure in chemical equilibrium) can be expressed as a linear combination of the chemical imbalances 
\begin{equation}\label{eq.bulk_scalar_comb}
    \Pi= \Pi_1 \mu_1 + \Pi_2 \mu_2,
\end{equation}
we obtain that a Burgers-type equation describes the evolution with time of the bulk viscous pressure~\cite{Gavassino_2023,Hernandez:2025zxw}. The coefficients $\Pi_1$ and $\Pi_2$ are obtained following the procedure described in Appendix~\ref{sec.thermodynamics}. 

As a result, the time evolution of the bulk scalar pressure in nuclear matter with electrons and muons is given by the following Burgers equation 
\begin{equation}\label{eq.Burgers_equation}
\det \tau\, D_u^2\Pi + \text{Tr}\, \tau \, D_u\Pi + \Pi =-\zeta \theta -\xi D_u\theta,
\end{equation}
where $D_u \equiv u^\mu \nabla_\mu$ is the convective derivative. The second-order transport coefficients $\zeta$ and $\xi$ are given by 
\begin{equation}
\begin{aligned}
    \zeta &\equiv \Pi^\alpha b_\alpha,\\
    \xi &\equiv \det\tau \, \Pi^\alpha \left(\tau^{-1} \right)_\alpha^{\,\,\beta} b_\beta,    
\end{aligned}
\label{eq.2ndordtransport}
\end{equation}
with $\Pi_\alpha\equiv(\Pi_1, \Pi_2)$.

The Green's function of the Burgers equation in the fluid rest frame, $u^\mu= (1,0,0,0)$, can be written down as the sum of two Green's function of a Israel-Stewart equation 
\cite{Gavassino_2023}
\begin{equation}\label{eq.Greenfunction}
    G(t)=G_+(t)+G_-(t),
\end{equation}
with
\begin{equation}\label{eq.Greenfunctionpm}
    G_{\pm}(t)\equiv \frac{\zeta_{\pm}}{\tau_{\pm}}\Theta_H(t)e^{-t/\tau_{\pm}},
\end{equation}
where $\tau_\pm$ are the eigenvalues of $\tau$,  
and $\zeta_\pm$ are the bulk viscosity components which are related with the transport coefficients as follows
\begin{equation}\label{eq.transportcoeff}
\begin{aligned}
    \zeta&=\zeta_+ +\zeta_-,\\
    \xi&=\zeta_+ \tau_- +\zeta_- \tau_+.    
\end{aligned}
\end{equation}
The parameter $\zeta$ is the total viscosity of the system. We will compute in Sec.~\ref{sec.results} and plot the partial viscosities, keeping in mind
that the total viscosity is provided by the sum of the components.

A solution to Eq.~\eqref{eq.Burgers_equation} can be written as 
\begin{equation}
    \Pi(t)=-\int_{-\infty}^\infty G(t') \nabla_\mu u^\mu (t-t')dt'.
\end{equation}

Explicit expressions for the whole set of  second-order coefficients %$(\tau_+,\zeta_+,\tau_-,\zeta_-)$ 
in terms of the scattering rates and susceptibilities are provided in Appendix \ref{app.transportcoefficients}.

\subsection{Frequency-dependent bulk viscosity} 
\label{sec-freqbulk}

In different astrophysical scenarios it is common to study the dissipation occurring when there is a perturbation periodic in time, characterized by a frequency $\omega$. From first-order
hydrodynamics one can derive a frequency-dependent bulk viscosity. However, it can also be recovered using the formalism derived in Sec.~\ref{section:Burgers_eq}~\cite{Gavassino_2023}. 
Let us see how this is achieved.

First, consider the Burgers equation Eq.~\eqref{eq.Burgers_equation} in the rest frame, that is,  using the fluid-velocity  $u_{\rm LRF}^\mu=(1,0,0,0)$, 
and  assume perturbations periodic in time, such that  $\Pi \propto e^{-i\omega t}$ and $\theta \propto e^{-i\omega t}$. Then the bulk scalar can be obtained as
\begin{equation}\label{eq.bulkscalar}
    \Pi =-\frac{\zeta-i\omega \xi}{1-i\omega \text{Tr}\,\tau - \omega^2 \det \tau} \, \theta \ .
\end{equation}
The frequency-dependent bulk viscosity is obtained in first order hydrodynamics from the real part of the bulk viscous pressure 
\begin{equation}
    \zeta(\omega)\equiv-\frac{\text{Re}[\Pi]}{\theta}.
\end{equation}
Thus, the frequency-dependent bulk viscosity is expressed as
\begin{equation}\label{eq.bulk_viscosity_eff}
    \zeta(\omega) =\frac{\zeta/(\det \tau)^2 +\omega^2(\xi \text{Tr}\,\tau-\zeta\det \tau)/(\det \tau)^2}{(1/\det \tau -\omega^2 )^2 +\omega^2 (\text{Tr}\,\tau/\det \tau)^2}.
\end{equation}

Moreover,  we note that the bulk scalar in Eq.~\eqref{eq.bulkscalar} can also be written as 
\begin{equation}
    \Pi=\left(\frac{\zeta_+}{i\omega \tau_+ -1} + \frac{\zeta_-}{i\omega \tau_- -1} \right) \theta,
\end{equation}
and this expression enables us to split the frequency-dependent bulk viscosity into two contributions as follows
\begin{equation} \label{eq.bulk_viscosity_eff_new}
    \zeta(\omega)=\frac{\zeta_+}{1+\omega^2 \tau_+^2} +\frac{\zeta_-}{1+\omega^2 \tau_-^2} \ ,
\end{equation}
in concordance to Eq.~(\ref{eq.Greenfunction}), which expresses the fact that the solution to the
Burgers equation can be represented as the sum of two solutions of an Israel-Stewart equation.
Note that the values obtained for $\tau_+$ and $\tau_-$ (see Appendix~\ref{app.transportcoefficients})  cannot be interpreted in general as being associated exclusively to one of the two reaction channels $\lambda_1$ and $\lambda_2$. This only happens in the high frequency limit \cite{Harris:2024evy}.
It is also important to note that if we set $\lambda_2 =0$, when muons densities would be frozen, there is only one relaxation time $\tau_+$, while $1/ \tau_{-} =0$ (see Appendix~\ref{sec-app-peak}). 

We note that Eq.(\ref{eq.bulk_viscosity_eff_new}) exhibits in general two maximal values when plotted as a function of the temperature. We have not identified analytically where these resonant
peaks are, but our numerical results suggest that they occur  when $\omega \approx 1/\tau_+$ and $\omega \approx 1/\tau_-$.
We also find that $\tau_+ \approx \tau_-$ when $\lambda_1 \approx \lambda_2$, and there is only one resonant peak. This is the situation which
is predicted by many models of EoS of nuclear matter with muons \cite{Alford_2023}. It is also the situation when there are not muons. We discuss in the remaining part of this manuscript different scenarios where one or two resonances are present in the temperature profile of the bulk viscosity and its strong dependence on the nuclear parameters in Eq.~\eqref{eq.energy_density}, and thus on the dUrca density thresholds, which are different for electrons and for muons.

\section{Numerical Results for transport Coefficients}\label{sec.results}

In this section we present numerical results for the different second-order transport coefficients, as
well as for the frequency-dependent bulk viscosity, as a function of the parameters used in the metamodel  EoS, and for different values of the temperature and density. We only focus on temperatures ranging from $0.5$ MeV up to $10$ MeV, the expected maximal value of the temperature for the neutrino-transparent regime~\cite{Alford:2018lhf, Roberts_2012}.
Although our expressions take into account the dependence of all nuclear parameters in Eq.~\eqref{eq.energy_density}, as stated above, in the remaining part of the manuscript we fix the values of the nuclear parameters as 
\begin{equation}
    J =32 \,{\rm MeV}, \quad K=240 \,{\rm MeV},
\end{equation}
 and set $ K_{\rm sym}= Q =Q_{\rm sym}=0 $, given their poor constraints, and consider three  values of the slope $L=50$ MeV, $L=70$ MeV and $L=110$ MeV.            
Additionally, we neglect $\lambda_3$ in our analysis because the muon decay rate is much slower than semileptonic processes in the temperature regime considered (see Fig. 8 of Ref.~\cite{Harris:2024evy}). However, our derived expressions in Appendix \ref{app.transportcoefficients} are fully general and also consider its contribution.

\begin{figure}[t]
\includegraphics[width=0.483\textwidth]{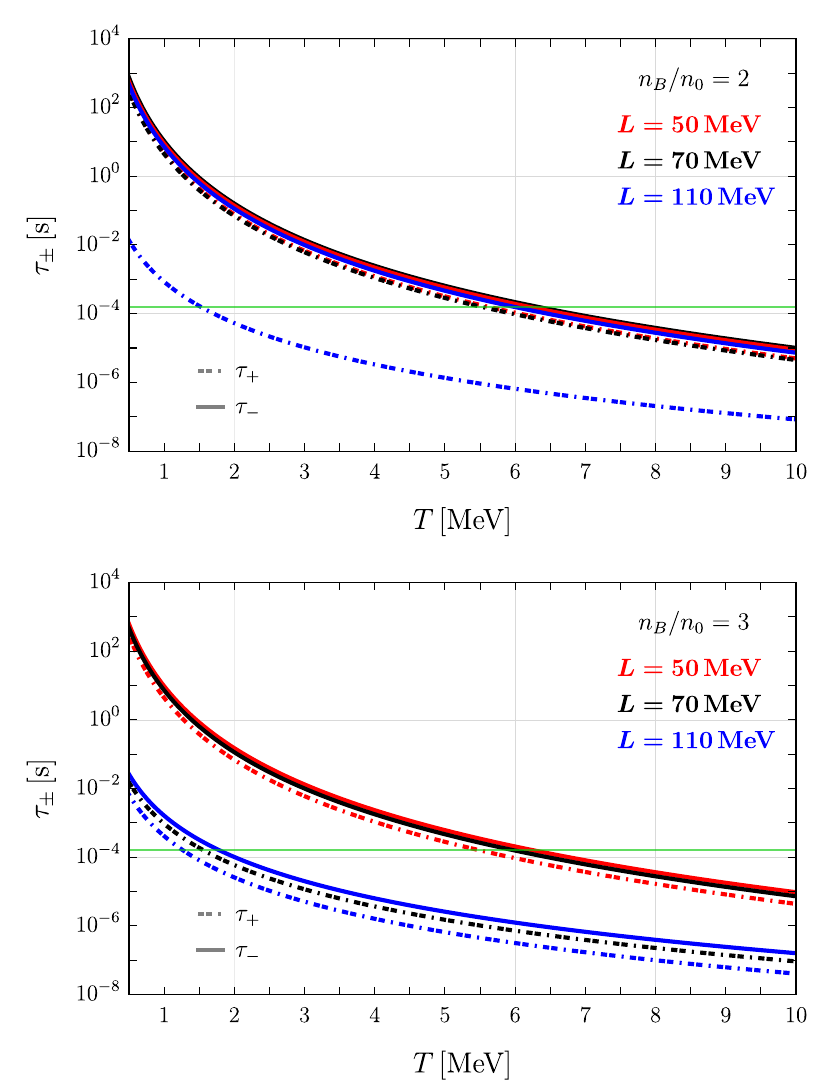}
\caption{Relaxation times $\tau_\pm$ as a function of temperature $T$ in log scale at $n_B=2 n_0$ (top panel) and $n_B=3 n_0$ (bottom panel); $\tau_+$ (dashed-dotted lines) and $\tau_-$ (continuous lines).  
The red, black and blue curves represent $L=50$ MeV, $L=70$ MeV and $L=110$ MeV, respectively. We also include a green line that represents $\omega^{-1}$ so that $2\pi/\omega=10^{-3}$ s}.\label{fig:taus_with_temperature_J32_K240_Ksym0_varying_L_and_density}
\end{figure}

In Fig.~\ref{fig:taus_with_temperature_J32_K240_Ksym0_varying_L_and_density}  we show the temperature profile of relaxation times in log scale, $\tau_+$ (dashed-dotted lines) and $\tau_-$ (continuous lines) and in Fig.~\ref{fig:zeta_with_temperature_J32_K240_Ksym0_varying_L_and_density} the temperature profile of bulk viscosity components in log scale, $\zeta_+$ (dashed-dotted lines) and $\zeta_-$ (continuous lines). For both figures we use the same baryon number densities and values of the slope of the symmetry energy. This is, we study $\tau_\pm$ and $\zeta_\pm$ at $n_B=2n_0$ (top panels) and $n_B=3 n_0$ (bottom panels) and $L=50$ MeV (red curves), $L= 70$ MeV (black curves) and $L=110$ MeV (blue curves). We also include a continuous green line at $\omega^{-1}$ where $2\pi/\omega=10^{-3}$ s (with $\omega$ a typical frequency in mergers) in Fig.~\ref{fig:taus_with_temperature_J32_K240_Ksym0_varying_L_and_density} to identify the temperatures where relaxation times match this inverse angular frequency. When dashed-dotted and continuous lines of the same color are on top of each other or very close, we only expect one resonant peak in the frequency bulk viscosity, while in the opposite case, we will expect two resonant peaks. For all the cases considered in Fig.~\ref{fig:taus_with_temperature_J32_K240_Ksym0_varying_L_and_density}, this latter trend occurs at $L=110$ MeV and $n_B=2n_0$ (blue curves in the top panel) and $L=70$ MeV and $n_B=3n_0$ (black curves in the bottom panel). These parameter sets provide the conditions for only dUrca processes with electrons to be kinematically allowed (see Fig.~\ref{fig:thrdU_with_slope_nuclear_matter_J32_K240_Ksym0}) producing a significant increase on $\lambda_1$ with respect to $\lambda_2$ as can be observed in Fig.~\ref{fig:temperature_lambdas_with_J32_K240_varying_L_and_density}.

\begin{figure}[t]
\includegraphics[width=0.483\textwidth]{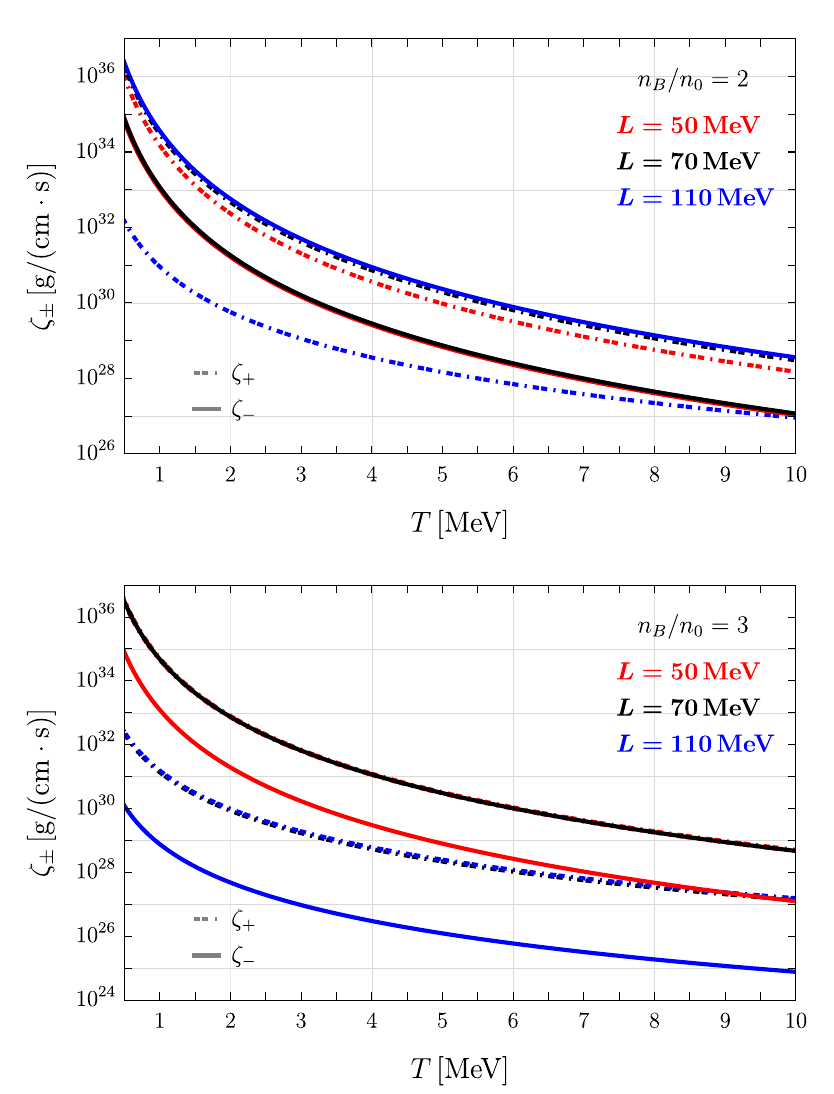}
\caption{Bulk viscosity components $\zeta_\pm$ as a function of temperature $T$ in log scale at $n_B=2 n_0$ (top panel) and $n_B=3 n_0$ (bottom panel); $\zeta_+$ (dashed-dotted lines) and $\zeta_-$ (continuous lines).  
%The different curves are obtained at $J=32$ MeV and $K=240$ MeV. 
The red, black and blue curves represent $L=50$ MeV, $L=70$ MeV and $L=110$ MeV, respectively.}
\label{fig:zeta_with_temperature_J32_K240_Ksym0_varying_L_and_density}
\end{figure}

For the parameter sets considered in this work, relaxation times and bulk viscosity components decrease significantly as temperature increases; we also observe a slight variation of these transport coefficients with respect to baryon number density unless is high enough at fixed $L$ to reach dUrca processes thresholds. For instance, at $L=50$ MeV (red curves in both panels of Figs.~\ref{fig:taus_with_temperature_J32_K240_Ksym0_varying_L_and_density} and \ref{fig:zeta_with_temperature_J32_K240_Ksym0_varying_L_and_density}) and $L=70$ MeV and $n_B=2n_0$ (black curves in the top panels of Figs.~\ref{fig:taus_with_temperature_J32_K240_Ksym0_varying_L_and_density} and \ref{fig:zeta_with_temperature_J32_K240_Ksym0_varying_L_and_density}), the chemical equilibration in the system is carried out by mUrca processes. However, at $L=110$ MeV and $n_B=3n_0$ (blue curves in the bottom panels of Figs.~\ref{fig:taus_with_temperature_J32_K240_Ksym0_varying_L_and_density} and \ref{fig:zeta_with_temperature_J32_K240_Ksym0_varying_L_and_density}), the slope and density are high enough to access to the kinematic region where both dUrca processes with electrons and muons are allowed (above the shaded gray region in  Fig.~\ref{fig:thrdU_with_slope_nuclear_matter_J32_K240_Ksym0}). As a result, we observe a significant decrease of second-order transport coefficients. Additionally, we consider the case where only dUrca processes with electrons are allowed (shaded gray region in Fig.~\ref{fig:thrdU_with_slope_nuclear_matter_J32_K240_Ksym0}), this occurs at $L=110$ and $n_B=2n_0$ (blue curves in the top panels of Figs.~\ref{fig:taus_with_temperature_J32_K240_Ksym0_varying_L_and_density} and \ref{fig:zeta_with_temperature_J32_K240_Ksym0_varying_L_and_density}) and at $L=70$ MeV and $n_B=3n_0$ (black curves in the bottom panels of Figs.~\ref{fig:taus_with_temperature_J32_K240_Ksym0_varying_L_and_density} and \ref{fig:zeta_with_temperature_J32_K240_Ksym0_varying_L_and_density}). The resulting behavior is the dominance of $\zeta_-$ over the $\zeta_+$, opposite to the trend obtained for the other cases mentioned before. For relaxation times, despite $\tau_->\tau_+$ in the temperature regime considered and for all the sets studied, we observe that the gap between them is enhanced significantly for the parameter sets where only electronic dUrca together with mUrca processes contribute to the chemical equilibration.

\begin{figure}[b]
\includegraphics[width=0.483\textwidth]{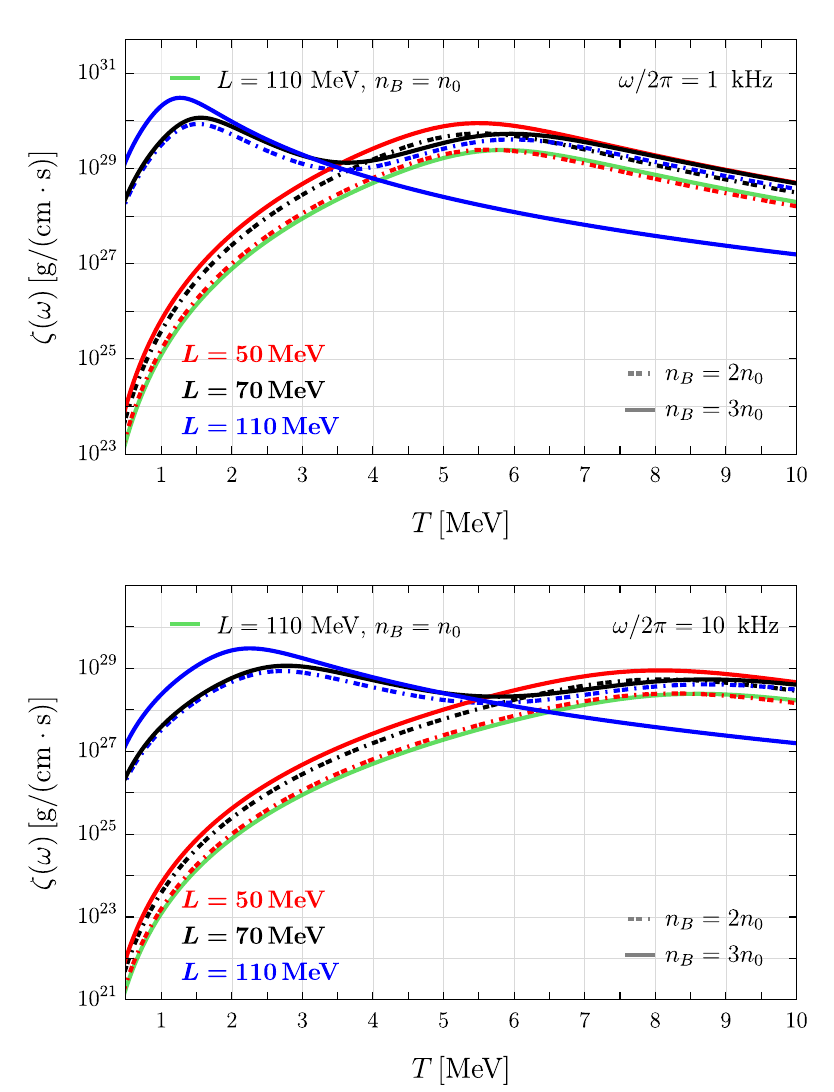}
\caption{Frequency-dependent bulk viscosity $\zeta(\omega)$ as a function of the temperature $T$ in log scale at $\omega/2\pi=1$ kHz (top panel) and $\omega/2\pi=10$ kHz (bottom panel) and $n_B=2 n_0$ (dashed-dotted lines) and $n_B=3n_0$ (continuous lines). 
%The different curves are obtained at $J=32$ MeV and $K=240$ MeV. 
The red, black and blue curves represent $L=50$, $L=70$ MeV and $L=110$ MeV, respectively. We also include a solid green line to display the case $n_B=n_0$ and $L=110$ MeV.}
\label{fig:effective_zeta_with_temperature_J32_K240_Ksym0_varying_L_and_density}
\end{figure}

In Fig.~\ref{fig:effective_zeta_with_temperature_J32_K240_Ksym0_varying_L_and_density} we plot the temperature profile of the frequency-dependent bulk viscosity at $\omega/2\pi=1$ kHz (top panel) and $\omega/2\pi=10$ kHz (bottom panel). 
We set $L=50$ MeV (red curves), $L=70$ MeV (black  curves), and $L=110$ MeV (blue curves) and also consider $n_B=2n_0$ (dashed-dotted lines) and $n_B=3n_0$ (continuous lines) for comparison with second-order transport coefficients in Figs.~\ref{fig:taus_with_temperature_J32_K240_Ksym0_varying_L_and_density} and \ref{fig:zeta_with_temperature_J32_K240_Ksym0_varying_L_and_density}. We also consider $n_B=n_0$ at $L=110$ (continuous green line).

We observe that for $L=50$ MeV and $\omega/2\pi=1$ kHz (red curves in the top panel), the bulk viscosity profile reaches its maximum at $T\approx 5.63$ MeV and $T\approx 5.50$ MeV when $n_B=2n_0$ and $n_B=3n_0$, respectively. And for $\omega/2\pi= 10$ kHz (red lines in the bottom panel) this maximum is shifted to $T\approx 8.26$ MeV and $T\approx 8.07$ MeV when $n_B=2n_0$ and $n_B=3n_0$, respectively. Particularly, this trend is in accordance with the resulting profile considering the microscopic DDME2 model in Ref.~\cite{Alford_2023},  whose value for $L= 51.27$ MeV \cite{Alford:2022ufz},  and $K= 251.5$ MeV, slightly above our value of $K= 240$ MeV, while the values of the nuclear parameters  $B_{\rm sat},\, n_0$ and $J$ are the same as those we used  and
$K_{\rm sym}$ is not specified. Note that for the DDME2 model, the density threshold for dUrca processes occurs above $5n_0$  \cite{Alford:2022ufz}.  With our choice of nuclear parameters, we can identify the density threshold for the opening of dUrca processes, which are at lower values of the density (see Fig.~\ref{fig:thrdU_with_slope_nuclear_matter_J32_K240_Ksym0}). We suspect that this could be due to the choice of $K_{\rm sym}$, which in this model is not specified. 
However, for the densities we consider in our plots, the viscosity is also dominated by mUrca, and we thus obtain  results in concordance with the DDME2 model.  

We  note that for  $L=110$ MeV and $n_B=3n_0$ (continuous blue curve in both panels) dUrca processes including electrons and muons are allowed and together with the mUrca processes slightly increase the maximum value of the bulk viscosity profile and shift it to lower temperatures at $T=1.26$ MeV  for $\omega/2\pi=1$ kHz and to $T\approx 2.25$ MeV for $\omega/2\pi=10 $ kHz.   This effect seems to be consistent with the results obtained from the microscopic NL3 model used in Ref.~\cite{Alford_2023}, with values of the nuclear parameters as 
$L= 118.9$ MeV,  $K= 251.5$ MeV, and  $J=37.4$ MeV, while the remaining parameters are the same as ours,    and $K_{\rm sym}$ is not specified \cite{Alford:2022ufz}. The threshold  densities for  electronic dUrca rates occurs at $n_B\approx1.3n_0$ and for muonic dUrca rates at $n_B\approx 1.6 n_0$,   while the threshold densities with the choice of parameters of Sec.~\ref{sec-EOS} are  slightly below those values. For the choice  $n_B= 3 n_0$ we find that the viscosity is then dominated by dUrca,  and we obtain for this density results on concordance with the results discussed for the NL3 model in \cite{Alford:2022ufz}.

We also observe an interesting behavior when 
only dUrca processes with electrons are allowed (shaded gray region in Fig.~\ref{fig:thrdU_with_slope_nuclear_matter_J32_K240_Ksym0}). For instance, at $L=110$ MeV and $n_B=2n_0$ (dashed-dotted blue lines in both panels) and at $L=70$ MeV and $n_B=3n_0$ (continuous black lines in both panels) for $\omega/2\pi=1$ kHz, we observe two local maxima at $T\approx 1.52$ MeV and $T\approx 5.96$ MeV and at $T\approx 1.56$ MeV and $T\approx 5.96$ MeV, respectively. For $\omega/2\pi=10$ kHz they are slightly shifted to higher temperatures. These local maxima seem to occur when $\tau_\pm^{-1} \approx \omega$, although it seems complicated to find analytical expressions for the $T$ when the maxima occurs in full generality (see Appendix~\ref{sec-app-peak}). 

It is also relevant to mention that this behavior is not necessarily restricted to high values of $L$. As we can see in the shaded region in Fig.~\ref{fig:thrdU_with_slope_nuclear_matter_J32_K240_Ksym0}, a similar scenario can be produced for much lower values of the slope as we slightly increase the baryon number density. For $L=50$ MeV this effect occurs but at very high densities.

Note that we have kept $K_{\rm sym}=0$ given its poorly constrained value, but we have studied  the variation of the dUrca density thresholds with $K_{\rm sym}$ to show the possible dependence of the threshold on its value in Fig.~\ref{fig.thrdU_with_curvature_nuclear_matter_J32_K240}. The range of values for $K_{\rm sym}$ in that figure was chosen such that the Taylor expansion performed in the metamodel is still valid. Within these reasonable limits for $K_{\rm sym}$, we find that for large values of $L$, the dUrca density onsets are almost independent of this value. Therefore, for models with similar $L$, such as NL3, we expect the double-peak to be robust against possible variations of $K_{\rm sym}$ within the validity region. As for smaller values of $L$, there is a stronger dependence with $K_{\rm sym}$ of the dUrca density thresholds, specially for models with $L \sim$ 50 MeV, such as DDME2. In that case, the two-double peak structure of the bulk viscosity will be correlated with $K_{\rm sym}$, always appearing once the dUrca for electrons is allowed and the one for muons forbidden, as indicated by the gray areas in Fig.~\ref{fig.thrdU_with_curvature_nuclear_matter_J32_K240}.

In Table~\ref{tab.table1} we show the local-maximum temperature in Fig.~\ref{fig:effective_zeta_with_temperature_J32_K240_Ksym0_varying_L_and_density} for the parameter sets where we obtained a one-resonance peak. A numerical analysis allow us to determine that in this scenario these local-maximum temperatures approximately correspond to the values where $\tau_+=\omega^{-1}$. Moreover, in Table~\ref{tab.table2} we display the two local maxima and the intermediate minimum in a two-resonance peak behavior of the temperature profile of frequency-dependent bulk viscosity in Fig.~\ref{fig:effective_zeta_with_temperature_J32_K240_Ksym0_varying_L_and_density}. In both tables, we note that the local maxima and minimum shift to higher temperatures as we increase the angular frequency up to $\omega/2\pi=10$ kHz. Particularly, in Table~\ref{tab.table2} we observe small variation of the local maxima and minimum with the slope and baryon number density. 

\begin{table}[H] 
  \centering
  \begin{tabular}{| c | c | c | c |}
    \hline
    $T_{\rm max}$  [MeV]&  $L$ [MeV] & $n_B/n_0$ & $\omega/2\pi$ [kHz] \\
    \hline
    5.63                & 50  & 2 & 1  \\ 
    8.26                & 50  & 2 & 10 \\ 
    5.50                & 50  & 3 & 1  \\ 
    8.07                & 50  & 3 & 10 \\ 
    5.52                & 70  & 2 & 1  \\ 
    8.10                & 70  & 2 & 10 \\ 
    1.26                & 110 & 3 & 1  \\ 
    2.25                & 110 & 3 & 10 \\
    5.85                & 110 & 1 & 1  \\ 
    8.59                & 110 & 1 & 10 \\ 
    \hline
  \end{tabular}
  \caption{Local maximum of the  temperature profile of frequency-dependent bulk viscosity for the sets where there is only one-resonance peak.}
\label{tab.table1}
\end{table}

\begin{table}[H] 
  \centering
  \begin{tabular}{| c | c | c | c | c |}
    \hline
    $T_{\rm max}$  [MeV]& $T_{\rm min}$ [MeV] & $L$ [MeV] & $n_B/n_0$ & $\omega/2\pi$ [kHz] \\
    \hline
    1.56, 5.96       & 3.63 & 70  & 3 & 1  \\ 
    2.77, 8.72       & 5.76 & 70  & 3 & 10 \\ 
    1.52, 5.96       & 3.58 & 110 & 2 & 1  \\ 
    2.70,  8.73       & 5.68 & 110 & 2 & 10 \\ 
    \hline
  \end{tabular}
  \caption{Local maxima the intermediate minimum temperature in a two-resonance peak behavior of the temperature profile of frequency-dependent bulk viscosity for the parameter sets considered in this work.}
\label{tab.table2}
\end{table}

For completeness, we study second-order transport coefficients and frequency-dependent bulk viscosity as a function of baryon number density. In Fig.~\ref{fig:tau_with_density_J32_K240_Ksym0_varying_L_and_temperature}, we show the relaxation times $\tau_+$ (dashed-dotted curves) and $\tau_-$ (continuous curves) and in Fig.~\ref{fig:zetas_with_density_J32_K240_Ksym0_varying_L}, the bulk viscosity components, $\zeta_+$ (dashed-dotted lines) and $\zeta_-$ (continuous lines). For both figures we set $T=1$ MeV and consider $L=50$ MeV (red lines), $L=70$ MeV (black lines), and $L=110$ MeV (blue lines).

\begin{figure}[b]
\includegraphics[width=0.483\textwidth]{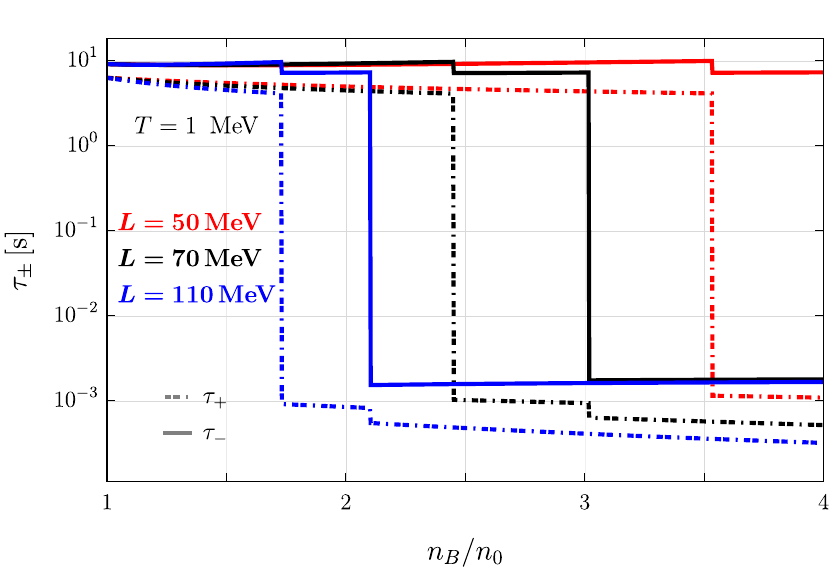}
\caption{Relaxation times $\tau_\pm$ as a function of baryon number density $n_B$ in multiples of the saturation density $n_0$ in log scale at $T=1$ MeV; $\tau_+$ (dashed-dotted lines) and $\tau_-$ (continuous lines). The red curves represent $L=50$ MeV, while black and blue lines  to $L= 70$ and  $L=110$ MeV, respectively.}
\label{fig:tau_with_density_J32_K240_Ksym0_varying_L_and_temperature}
\end{figure}

In Fig.~\ref{fig:tau_with_density_J32_K240_Ksym0_varying_L_and_temperature} we note, despite the relaxation times involve both rates $\lambda_1$ and $\lambda_2$, it is possible to infer their individual impact on $\tau_+$ and $\tau_-$ when baryon number density reaches the density threshold of dUrca processes. At these thresholds, dUrca processes produce a piecewise curve of relaxation times with two visible decreases. On one hand, $\tau_+$ suffers an abrupt change of around four orders of magnitude due to the emergence of dUrca processes with electrons, while a second step much smaller can be observed when dUrca processes with muons are allowed. On the other hand, $\tau_-$ shows an opposite trend with a sharp decrease when dUrca processes with muons emerge. Typically, $\tau_-$ is slightly higher than $\tau_+$, but in the density regime when only dUrca processes with electrons are allowed at fixed $L$, $\tau_-$ is around four orders of magnitude higher than $\tau_+$. 

\begin{figure}[t]
\includegraphics[width=0.483\textwidth]{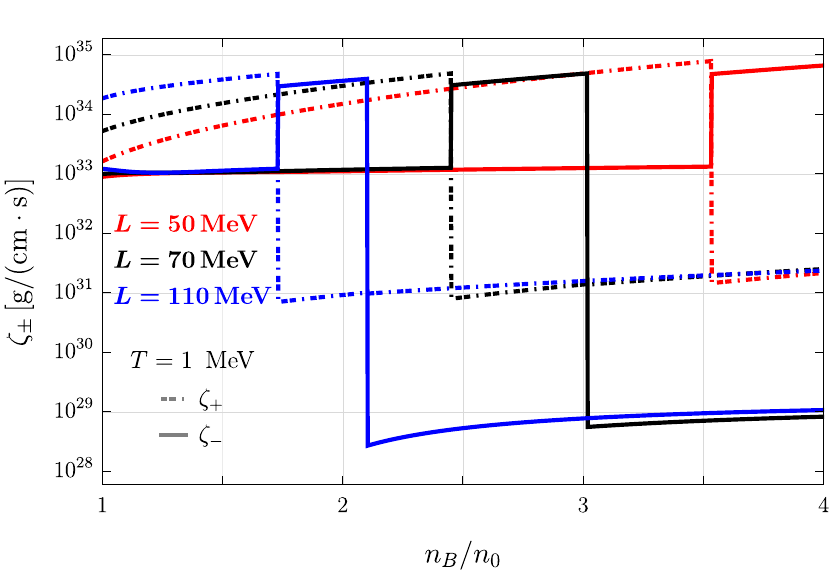}
\caption{Bulk viscosity components $\zeta_\pm$ as a function of baryon number density $n_B$ in multiples of the saturation density $n_0$ in log scale at $T=1$ MeV; $\zeta_+$ (dashed-dotted lines) and $\zeta_-$ (continuous lines). The red curves represent $L=50$ MeV, while black and blue lines  to $L= 70$ and  $L=110$ MeV, respectively.}
\label{fig:zetas_with_density_J32_K240_Ksym0_varying_L}
\end{figure}

In Fig.~\ref{fig:zetas_with_density_J32_K240_Ksym0_varying_L}, for all the cases considered, bulk viscosity components are also a piecewise function of baryon density with the domain partitioned as dUrca processes open. In particular, $\zeta_+$ is an increasing function of baryon number density before the density threshold of dUrca processes with electrons. Slightly above this value, $\zeta_+$ decreases around four orders of magnitude and then it continues increasing with $n_B$. Note this component does not experience a visible change when muonic dUrca processes emerge. Moreover, $\zeta_-$ changes abruptly its almost constant trend when dUrca thresholds are reached. First, dUrca processes with electrons generate an increase slightly above one order of magnitude, while the emergence of dUrca processes including muons produce an abrupt decrease of almost seven orders of magnitude.  
Recall that the total (frequency-independent) bulk viscosity $\zeta$ is given by the sum of the two components, and thus also $\zeta$ changes by orders of magnitude for some fixed values of $n_B$ when $L$ is changed. 

In Fig.~\ref{fig:effective_zeta_with_density_J32_K240_Ksym0_varying_L}, we consider the frequency-dependent bulk viscosity as a function of the baryon number density in multiples of the saturation density $n_0$ at $\omega/2 \pi=1$ kHz. We set $T=1$ MeV (dotted-dashed curves) and $T=5$ MeV (continuous curves) and vary the slope of the symmetry energy for $L=50$ MeV, $L=70$ MeV and $L=110$ MeV in red, black and blue lines, respectively. Note that dUrca processes have different effects on the frequency-dependent bulk viscosity for the two different temperatures considered. This is, when electronic dUrca processes contribute to the chemical equilibration, at $T=1$ MeV, they produce a sudden increase in bulk viscosity around four orders of magnitude, while at $T=5$ MeV the effect is opposite, making the bulk viscosity almost one order of magnitude smaller. Furthermore, when density is high enough for the emergence of dUrca processes with muons, at $T=1$ MeV this contribution produce a slight increase of bulk viscosity, while at $T=5$ MeV the bulk viscosity suffer a moderate decrease slightly above one order of magnitude. 
The slope only varies the density threshold of dUrca processes and does not seem to impact on how bulk viscosity changes. Our results at $T=1$ MeV are in qualitative agreement with previous studies~\cite{Haensel:1992zz,Haensel_2001} where $T=10^9 \,\text{K}\approx0.1$ MeV and $\omega=10$ kHz are considered. The reported behavior also suggests that there is an intermediate temperature where the emergence of dUrca processes with electrons or muons do not produce a noticeable change on the bulk viscosity. In fact, this can be observed in Fig.~\ref{fig:effective_zeta_with_temperature_J32_K240_Ksym0_varying_L_and_density}. For dUrca processes with electrons this occurs at the temperature where the black curves at $L=70$ MeV cross. As stated above, at $n_B=2n_0$ (dashed-dotted black curves), the equilibration is leading by mUrca processes, but at $n_B=3n_0$ (solid black curves), dUrca processes with electrons are allowed. For dUrca processes with muons, we infer this temperature by identifying the temperature where blue curves at $L=110$ MeV cross each other.

\begin{figure}[b]
\includegraphics[width=0.483\textwidth]{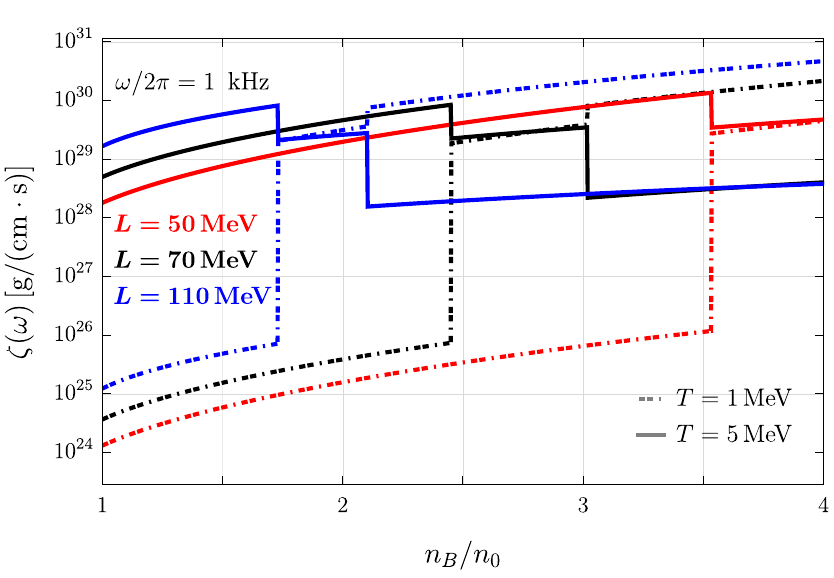}
\caption{Frequency-dependent bulk viscosity $\zeta(\omega)$ as a function of the baryon density in multiples of the saturation density $n_0$ in log scale at $\omega/2\pi=1$ kHz and $T=1$ MeV (dashed lines) and $T=5$ MeV (continuous lines). 
The red curves represent $L=50$ MeV, while black and blue lines  to $L= 70$ and  $L=110$ MeV, respectively.}
\label{fig:effective_zeta_with_density_J32_K240_Ksym0_varying_L}
\end{figure}

\section{Damping times of density oscillations}
\label{sec-damping}

In this section we determine the damping time associated to the dissipative bulk viscosity in the presence of density oscillations of frequency $\omega$, which can be obtained after evaluating~\cite{Sawyer:1989dp,Alford:2017rxf}

\begin{equation}
    \tau_\zeta = \frac{n_{B}^2}{\omega^2 \zeta (\omega)} \frac{\partial^2 \varepsilon}{\partial n_B^2}\bigg|_{X_p,X_e} ,
\end{equation}
and thus, depends on the EoS used to model the neutron star matter. In particular, with the use of the metamodel, we obtain 
\begin{equation}
    \begin{split}
        \frac{\partial^2 \varepsilon}{\partial n_B^2}\bigg|_{X_p,X_e}&=\frac{3n_B-2n_0}{9n_0^2}K+\frac{(n_B-n_0)(2n_B-n_0)}{27n_0^3}Q \\
        &\left(\frac{\pi X_e^2}{3n_B}\right)^{2/3}+\beta^2\left[\frac{2}{3n_0}L+\frac{(3n_B-2n_0)}{9n_0^2} K_{\rm sym} \right.\\
        &\left. +\frac{(n_B-n_0)(2n_B-n_0)}{27n_0^3}Q_{\rm sym} \right] \\
        &+\frac{\pi^{4/3} (X_p-X_e)^{5/3}}{(3n_B)^{1/3}\sqrt{m_\mu^2+[3\pi^2n_B(X_p-X_e)]^{2/3}}}.
    \end{split}
\end{equation}
for electrically neutral matter.

\begin{figure}[t]
\includegraphics[width=0.483\textwidth]{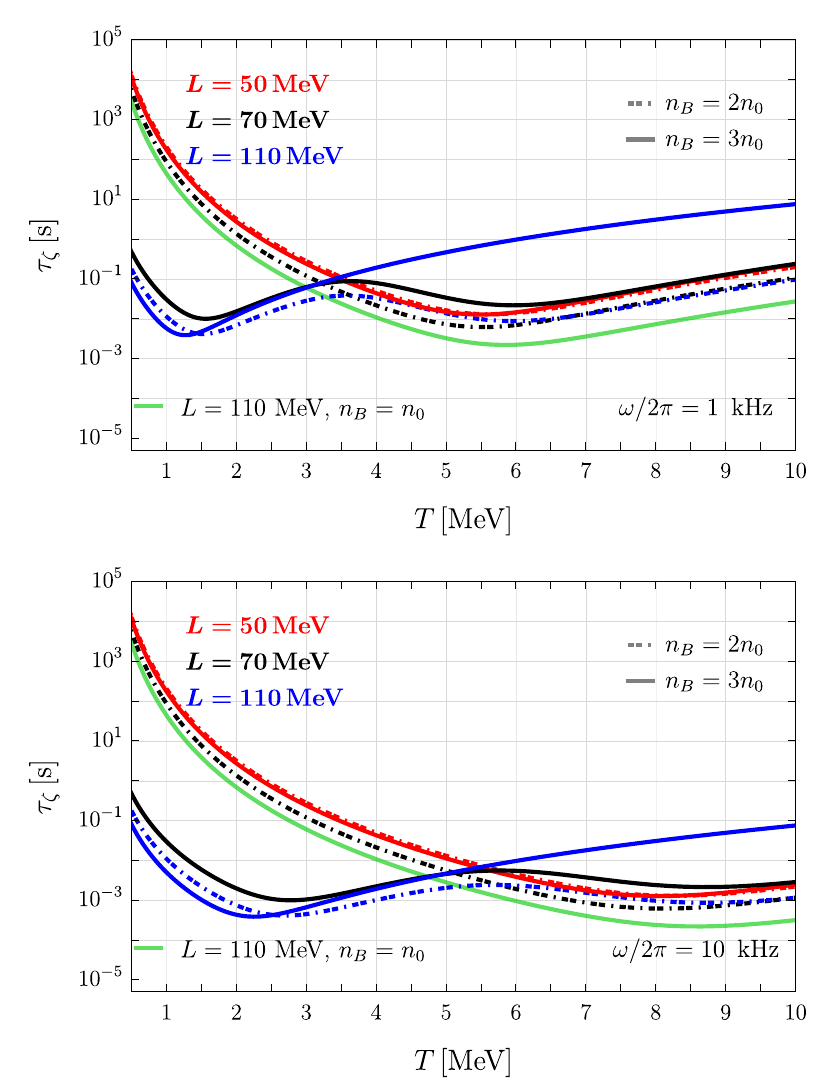}
\caption{Damping times $\tau_\zeta$ as a function of the temperature $T$ in log scale at $\omega/2\pi=1$ kHz (top panel) and $\omega/2\pi=10$ kHz (bottom panel)  and $n_B=2 n_0$ (dashed lines) and $n_B=3n_0$ (continuous lines). The red, black and blue curves represent $L=50$, $L=70$ MeV and $L=110$ MeV, respectively. We also include a solid green line to display the case $n_B=n_0$ and $L=110$ MeV.}
\label{fig:DAMPING_J32_K240_Ksym0_varying_L_and_density}
\end{figure}

In Fig.~\ref{fig:DAMPING_J32_K240_Ksym0_varying_L_and_density}, we report the corresponding values of the damping times for the cases studied in our temperature profiles of the frequency-dependent bulk viscosity in Fig.~\ref{fig:effective_zeta_with_temperature_J32_K240_Ksym0_varying_L_and_density}. First, we can note that damping times significantly decrease with angular frequency. For all the cases reported, at $\omega/2\pi=1$ kHz damping times are at least one order of magnitude higher than the ones obtained at $\omega/2\pi=10$ kHz where damping times are below the millisecond scale ($L=110$ MeV at $n_B/n_0=1,\,2,\,3$). Note that the lowest values of the damping times obtained are not necessarily correlated with the sets where bulk viscosity is the strongest one. Particularly at $n_B=n_0$ and $L=110$ MeV (continuous green line) produces the lowest damping times reported in Fig.~\ref{fig:DAMPING_J32_K240_Ksym0_varying_L_and_density} and the maximum value produced in its bulk viscosity is one of the lowest ones considered in Fig.~\ref{fig:effective_zeta_with_temperature_J32_K240_Ksym0_varying_L_and_density}. This effect is mainly associated to role of the incompressibility of nuclear matter
\begin{equation}
    \kappa =9n_B \frac{\partial^2 \varepsilon}{\partial n_B^2}\bigg|_{X_p,X_e},
\end{equation}
which acts like a spring constant of dense nuclear matter. The trend reported in Fig.~\ref{fig:DAMPING_J32_K240_Ksym0_varying_L_and_density} is also consistent with the results in Figs.[12,15,17,18] of Ref.~\cite{Alford_2023}.

Our results allow us to identity for which values of the different set nuclear parameters and densities we may expect to have the fastest damping of density oscillations, ranging from a few milliseconds to hundred milliseconds at $\omega/2\pi=1$ kHz for the densities and temperatures considered,  so as to evaluate whether the bulk viscosity might have an impact on the dynamics of neutron star mergers.

\section{Conclusions}
\label{sec-conclusions}

In this work we have set a formalism to study the bulk viscosity when there are two out-of-equilibrium chemical imbalances in the most possible general way, relating both the second-order transport coefficients that appear in the Burgers equation with the most commonly used frequency-dependent bulk viscosity. 
Then we have studied the bulk viscosity in the neutrino-transparent regime of neutron star matter  composed by neutrons, protons, electrons and muons using a metamodel to describe its thermodynamics.  There is  hope that the different nuclear parameters might be constrained, not only with terrestrial experiments, but also with different future astrophysical observations (see as possible examples 
\cite{Pradhan:2023zor,Sotani:2026wrk}). Fixing the nuclear parameters would allow to fix  the value of the viscosity. As emphasized in Ref.\cite{Harris:2025ncu},  there is little linkage in most part of the literature between computations of EoS and transport coefficients for NS. 

Our work should be viewed as an effort in the direction to provide a unified microscopic description  of neutron star physics, and could be readily implemented in numerical simulations of merger dynamics \cite{Most:2021ktk}. One should however keep in mind that the metamodel is only expected to be valid for densities not far away from the nuclear saturation density, while in mergers much higher values might be reached, which would then require a different modelization.  
As in the postmerger phase temperatures much larger than 10 MeV can be reached, and then neutrinos get trapped, our work should be generalized in that regime to describe those stages of the merger. Let us here also stress that there are very few simulations that include muons, and these studies conclude that their effect is not negligible \cite{Pajkos:2024iry,Gieg:2024jxs,Ng_2025}. 

Our results are consistent with those of Refs.~\cite{Alford_2023,Alford:2022ufz},  which were carried for two different microscopic models, when  some values of the nuclear parameters are similar, and find that the presence of muons slightly increase the value of the viscosity for fixed values of the different parameters.
However,  we found new effects after exploring the dependence of the viscosity on the slope of the symmetry energy. We have studied the density threshold for the opening of dUrca processes for electrons and muons as a function of the parameters $L$ and $K_{\rm sym}$ that describe the nuclear symmetry energy. We have focused our study on $L$, as  $K_{\rm sym}$ is poorly constrained. 
However, our Fig.~\ref{fig.thrdU_with_curvature_nuclear_matter_J32_K240} suggests that it would also be interesting to explore its effect. 
For every value  of $L$ there is a window of densities (those in the shaded gray region of 
Fig.~\ref{fig:thrdU_with_slope_nuclear_matter_J32_K240_Ksym0}) where we might expect a visible two-resonant peak behavior of the frequency-dependent bulk viscosity.  This is an effect entirely due to the presence of muons, and absent when nuclear matter is only composed of neutrons, protons and electrons, and which has not been found out before.
For low values of $L$ this effect is only noticeable at very high densities, but for
higher values of $L$ it appears at  values which are realized in the core of a NS.  
The effect changes the bulk viscosity by several orders of magnitude different from what one could naively expect, and should lead to relevant effects in the damping of density oscillations and of hydrodynamical fluctuating modes, that we hope to explore in a near future.

\section{Acknowledgments}
J.L.H. thanks the warm hospitality of the University of Osaka, where this work was partially completed, and Prof. Luca Baiotti and Yongjia Huang for interesting comments about this study.
We thank S. Säppi for useful discussions.
This work was also partly supported by the Spanish program Unidad de Excelencia María de Maeztu CEX2020-001058-M, financed by MCIN/AEI/10.13039/501100011033, and by the MaX-CSIC Excellence Award MaX4-SOMMA-ICE.
We also acknowledge support from the
project PID2022-139427NB-I00 financed by the Spanish MCIN/AEI/10.13039/501100011033/FEDER, UE (FSE+), as well as from the Generalitat de Catalunya under contract 2021 SGR 171. L.T. was also supported by the Grant CIPROM 2023/59 of Generalitat Valenciana. This work makes use of \texttt{SciPy} \cite{Virtanen:2019joe}, \texttt{jupyter} \cite{jupyter}, and \texttt{Mathematica}~\cite{Mathematica} software packages.

\section{Data Availability} 
The data reported in this work are openly available in the research repository Zenodo~\cite{data}.

\appendix
\section{Thermodynamics of nuclear matter including electrons and muons with a metamodel EoS}

\label{sec.thermodynamics}
Using the energy density in Eq.~\eqref{eq.energy_density} we are able to compute chemical potentials and susceptibilities at zero temperature in terms of the nuclear parameters mentioned above. 
The chemical potential of a particle species and inverse susceptibilities are given by the thermodynamic relations:
\begin{equation}\label{eq.chemical_potential}
    \mu_a=\frac{\partial \varepsilon}{\partial n_a}\bigg|_{s,n_b\neq n_a }\ , \quad (\chi^{-1})^b_a =\frac{\partial \mu_a}{\partial n_b}\bigg|_{s, n_c\neq n_b},
\end{equation}
with $a,b,c=n,\,p,\,e,\,\mu$ and $s$ the entropy density.

Using Eqs.~\eqref{eq.energy_density} and \eqref{eq.chemical_potential}, we have

\begin{equation}
\begin{split}
\mu_n
&= m-B_{\rm sat}
 + \frac{K}{2}x^2
 + \frac{Q}{6}x^3
 + S(n_B)\,\beta(2-\beta) \\
&+ \frac{n_B}{3n_0}
 \left[Kx + \frac{Q}{2}x^2
 + \beta^2\left(L + K_{\rm sym}x + \frac{Q_{\rm sym}}{2}x^2
 \right)\right],\\
\mu_p
&= m-B_{\rm sat}
 + \frac{K}{2}x^2
 + \frac{Q}{6}x^3
 - S(n_B)\,\beta(2+\beta) \\
&+ \frac{n_B}{3n_0}\left[Kx+\frac{Q}{2}x^2
 + \beta^2\left(L + K_{\rm sym}x + \frac{Q_{\rm sym}}{2}x^2\right)\right].
\end{split}
\end{equation}

For the susceptibilities, we obtain: 
\begin{equation}
(\chi^{-1})_e^e=\left(\frac{\pi}{3n_e}\right)^{2/3},
\end{equation}

\begin{equation}
(\chi^{-1})_\mu^\mu=\frac{\pi^{4/3}}{(3n_\mu)^{1/3}\sqrt{(3\pi^2 n_\mu)^{2/3}+m_\mu^2}},
\end{equation}

\begin{equation}
\begin{split}
(\chi^{-1})_n^{\,n}
&= \frac{2}{3n_0}
\left(Kx + \frac{Q}{2}x^2\right)
+ \frac{2}{n_B}(1-\beta)^2 S(n_B) \\
&
+ \frac{2}{3n_0}\,\beta
\left[1 + \frac{n_B}{n_0}(1-\beta)\right]
 \\
&\times \left(L + K_{\rm sym}x + \frac{Q_{\rm sym}}{2}x^2\right)\\
&
+ \frac{n_B}{9n_0^2}
\left[K + Qx + \beta^2\left(K_{\rm sym} + Q_{\rm sym}x\right)
\right],
\\
(\chi^{-1})_p^{\,p}
&= \frac{2}{3n_0}
\left(Kx + \frac{Q}{2}x^2\right)
+ \frac{2}{n_B}(1+\beta)^2 S(n_B) \\
&
- \frac{2}{3n_0}\,
\beta(2+\beta)
\left(L + K_{\rm sym}x + \frac{Q_{\rm sym}}{2}x^2\right) \\
&
+ \frac{n_B}{9n_0^2}
\left[K + Qx + \beta^2\left(K_{\rm sym} + Q_{\rm sym}x
\right)
\right].
\end{split}
\end{equation}
and the crossed-susceptibilities 
\begin{equation}
\label{crossed-susc}
\begin{aligned}
(\chi^{-1})_p^{\,n}
&= \frac{2}{3n_0}
\left(Kx + \frac{Q}{2}x^2\right)
- \frac{2}{n_B}(1-\beta^2)\,S(n_B) \\
&- \frac{2}{3n_0}\,
\beta^2
\left(L + K_{\rm sym}x + \frac{Q_{\rm sym}}{2}x^2
\right) \\
&+ \frac{n_B}{9n_0^2}
\left[K + Qx + \beta^2
\left(K_{\rm sym} + Q_{\rm sym}x\right)
\right].
\end{aligned}
\end{equation}
with $\left(\chi^{-1}\right)_p^{\,n}=\left(\chi^{-1}\right)_n^{\,p}$ and the rest of them are zero.

Furthermore, the total pressure of the system can be expressed as follows at linear order in deviations of the chemical potentials
\begin{equation}
    P=P_0+\sum_{a}\frac{\partial P}{\partial \mu_a}\bigg|_{\mu_b \neq \mu_a,T,\delta\mu_a=0} \delta \mu_a,
    \label{eq.bulkscalarinitial}
\end{equation}
where $P_0$ is the pressure in chemical equilibrium and the second term represents the out-of-equilibrium correction, that is the bulk scalar $\Pi$
\begin{equation}\label{eq.bulk_scalar}
    \Pi=n_n^{(\rm eq)}    
    \delta\mu_n +n_p^{(\rm eq)}\delta\mu_p+n_e^{(\rm eq)}\delta\mu_e+n_\mu^{(\rm eq)}\delta\mu_\mu.
\end{equation}
In what follows, we will drop the equilibrium superscript, to alleviate the notation. 
Initially, the four unknown deviations of the chemical potentials in Eq.~\eqref{eq.bulkscalarinitial} can be reduced to two by imposing constraints to keep the system local electric charge neutral and not to induce out-of-equilibrium deviations to the baryon number density, that is $\delta n_Q=0$ and $\delta n_B=0$. In this way, we get
\begin{equation}
\begin{split}
    \delta n_p &= \delta n_e +\delta n_\mu, \\
    \delta n_n &= -\delta n_e -\delta n_\mu, 
\end{split}
\end{equation}
or equivalently
\begin{equation}\label{eqs.constrains}
\begin{split}
    \chi_p^a \delta \mu_a &=(\chi_e^b +\chi_\mu^b)\delta \mu_b, \\
    \chi_n^a \delta \mu_a &=-(\chi_e^b +\chi_\mu^b)\delta \mu_b.
\end{split}
\end{equation}
From Eqs.~\eqref{eqs.constrains}, we get $\delta \mu_n$ and $\delta \mu_p$ as a function of $\delta \mu_e$ and $\delta \mu_\mu$.
\begin{equation}
    \begin{split}
        \delta \mu_n&=-\frac{\chi_n^{-1}\left[\delta \mu_\mu (\chi^{-1})_e^e+\delta \mu_e (\chi^{-1})_\mu^\mu\right]}{(\chi^{-1})_e^e (\chi^{-1})_\mu^\mu},\\
        \delta \mu_p&=\frac{\chi_p^{-1}\left[\delta \mu_\mu (\chi^{-1})_e^e+\delta \mu_e (\chi^{-1})_\mu^\mu\right]}{(\chi^{-1})_e^e (\chi^{-1})_\mu^\mu}.
    \end{split}
\end{equation}
To simplify the formulas, it is convenient to define the following shorthand
\begin{eqnarray}
    \chi_n^{-1}&\equiv& (\chi^{-1})_n^n-(\chi^{-1})_p^n \ ,\\
    \chi_p^{-1}&\equiv& (\chi^{-1})_p^p -(\chi^{-1})_p^n \ , \\
    \chi_1^{-1}&\equiv& (\chi^{-1})_e^e +\chi_n^{-1}+\chi_p^{-1} \ ,\\
    \chi_2^{-1}&\equiv& (\chi^{-1})_\mu^\mu +\chi_n^{-1}+\chi_p^{-1} \ , \\
    \chi_3^{-1}&\equiv&(\chi^{-1})_e^e+(\chi^{-1})_\mu^\mu \ . 
\end{eqnarray}
Note that we also write the resulting expressions in terms of the inverse susceptibilities obtained in this Appendix. This is done by employing the matrix of susceptibilities $\chi_a^b$ in terms of the inverse susceptibilities  
\begin{align}
\chi_a^b&\equiv\begin{pmatrix}(\chi^{-1})_p^p/D & -(\chi^{-1})_n^p/D & 
0 & 0\\
-(\chi^{-1})_p^n/D & (\chi^{-1})_n^n/D & 0 & 0\\
0 & 0 & 1/(\chi^{-1})_e^e & 0\\
0 & 0 & 0 & 1/(\chi^{-1})_\mu^\mu
\end{pmatrix},
\end{align}
with 
\begin{equation}
    D\equiv(\chi^{-1})_n^n (\chi^{-1})_p^p -(\chi^{-1})_n^p (\chi^{-1})_p^n,
\end{equation} 
the determinant of the submatrix that only contains the diagonal and crossed terms of nucleons in the inverse susceptibility matrix.

We use the definitions in Eqs.~\eqref{eq.chemicalimbalances} to get $\delta \mu_e$ and $\delta \mu_\mu$ as a linear combination of $\mu_1$ and $\mu_2$

\begin{equation}
\begin{split}
        \delta \mu_e &=-\frac{[(\chi_n^{-1} + \chi_p^{-1})(\mu_1 - \mu_2) + \mu_1  (\chi^{-1})_\mu^\mu](\chi^{-1})_e^e}{(\chi_n^{-1} + \chi_p^{-1})(\chi^{-1})_e^e + \chi_1^{-1}(\chi^{-1})_\mu^\mu}, \\
        \delta \mu_\mu &=\frac{[(\chi_n^{-1} + \chi_p^{-1})(\mu_1 - \mu_2) - \mu_2 (\chi^{-1})_e^e](\chi^{-1})_\mu^\mu}{(\chi_n^{-1} + \chi_p^{-1})(\chi^{-1})_e^e + \chi_1^{-1}(\chi^{-1})_\mu^\mu},
\end{split}
\end{equation}

Finally, using Eq.~\eqref{eq.bulk_scalar} we get Eq.~\eqref{eq.bulk_scalar_comb} 
with the coefficients $\Pi_1$ and $\Pi_2$ given by

\begin{equation}
  \begin{split}
    \Pi_1&=[-(\chi_n^{-1} + \chi_p^{-1})n_e(\chi^{-1})_e^e + (\chi_n^{-1}(n_n + n_\mu)  \\
    &+ \chi_p^{-1}(n_\mu-n_p) - n_e(\chi^{-1})_e^e)(\chi^{-1})_\mu^\mu]/ [(\chi_n^{-1} + \chi_p^{-1})\\
    &\times (\chi^{-1})_e^e  + \chi_1^{-1}(\chi^{-1})_\mu^\mu],\\
    \Pi_2&=[\chi_n^{-1}(n_e + n_n)(\chi^{-1})_e^e + \chi_p^{-1}(n_e - n_p)(\chi^{-1})_e^e \\
    &- n_\mu \chi_1^{-1}(\chi^{-1})_\mu^\mu]/ [(\chi_n^{-1} + \chi_p^{-1}) \\
    &\times(\chi^{-1})_e^e + \chi_1^{-1}(\chi^{-1})_\mu^\mu].
  \end{split}
\end{equation}

\section{Transport coefficients in nuclear matter} \label{app.transportcoefficients}
In this section, we provide the explicit expressions for the calculation of the transport coefficients in terms of susceptibilities, particle number densities and electroweak rates. First, the relaxation times are given by

\begin{equation}\label{eq.relaxationtimes}
\begin{split}
2\tau_{\pm}^{-1}
&=\pm \{(\lambda_1 \chi_1^{-1}+\lambda_2 \chi_2^{-1} +\lambda_3 \chi_3^{-1})^2 -4Q(\lambda_1,\lambda_2,\lambda_3) \\
&\times[(\chi^{-1})_\mu^\mu (\chi_n^{-1}+\chi_p^{-1})+(\chi^{-1})_e^e \chi_2^{-1}]\}^{1/2} \\
&+\lambda_1 \chi_1^{-1}
 + \lambda_2 \chi_2^{-1}
 + \lambda_3 \chi_3^{-1}.
\end{split}
\end{equation}
here we employ the sum of pairwise products $Q(x,y,x)\equiv xy+yz+zx$.

The bulk viscosity components have a complex and long expressions in terms of the combination of susceptibilities define above and particle number densities. Instead, we write $\zeta$ and $\xi$ and the bulk viscosity components can be obtained using the expressions Eqs.~\eqref{eq.transportcoeff} together with Eqs.\eqref{eq.relaxationtimes} 
\begin{equation}
\begin{split}
\zeta &= \frac{1}{Q(\lambda_1,\lambda_2,\lambda_3)
    [(\chi_n^{-1}+\chi_p^{-1})(\chi^{-1})_e^e+\chi_1^{-1}(\chi^{-1})_\mu^\mu]^2} \\
    &\times \Big\{ [\chi_n^{-1}(n_e + n_n)(\chi^{-1})_e^e + \chi_p^{-1}(n_e - n_p)(\chi^{-1})_e^e \\
    &- n_\mu \chi_1^{-1}(\chi^{-1})_\mu^\mu][(\chi_p^{-1} n_e \lambda_1 + \chi_n^{-1}(n_e + n_n)\lambda_1  \\
    &+ \chi_n^{-1} n_n \lambda_3- \chi_p^{-1} n_p(\lambda_1 + \lambda_3))(\chi^{-1})_e^e \\
    &- (\chi_n^{-1}(n_\mu \lambda_1 - n_n \lambda_3) + \chi_p^{-1}(n_\mu \lambda_1 + n_p \lambda_3) \\
    &+ n_\mu \lambda_1 (\chi^{-1})_e^e + (n_e + n_\mu)\lambda_3 (\chi^{-1})_e^e)(\chi^{-1})_\mu^\mu] \\
    &+[-(\chi_n^{-1} + \chi_p^{-1})n_e(\chi^{-1})_e^e + (\chi_n^{-1}(n_n + n_\mu) \\
    & + \chi_p^{-1}(-n_p + n_\mu)- n_e(\chi^{-1})_e^e)(\chi^{-1})_\mu^\mu] \\
    &\times [-(\chi_n^{-1} n_e \lambda_2 + \chi_p^{-1} n_e \lambda_2 - \chi_n^{-1} n_n \lambda_3 + \chi_p^{-1} n_p \lambda_3) \\
    &\times (\chi^{-1})_e^e + (\chi_p^{-1} n_\mu \lambda_2 + \chi_n^{-1}(n_n + n_\mu)\lambda_2  \\
    &+ \chi_n^{-1} n_n \lambda_3- \chi_p^{-1} n_p(\lambda_2 + \lambda_3)-(\chi^{-1})_e^e \\
    &\times (n_\mu \lambda_3 + n_e(\lambda_2 + \lambda_3)))(\chi^{-1})_\mu^\mu] \Big\} \ ,
\end{split}
\end{equation}
and 

\begin{equation}
\begin{split}
\xi &= \frac{1}{Q(\lambda_1,\lambda_2,\lambda_3)
    [(\chi_n^{-1}+\chi_p^{-1})(\chi^{-1})_e^e+\chi_1^{-1}(\chi^{-1})_\mu^\mu]^2} \\
    &\times \Big\{ (\chi^{-1})_e^e [(\chi_n^{-1} n_n - \chi_p^{-1} n_p)^2 + (\chi_n^{-1} + \chi_p^{-1})n_e^2 (\chi^{-1})_e^e]\\
    &+[(\chi_n^{-1} n_n - \chi_p^{-1} n_p)^2- 2(\chi_n^{-1} n_e n_n + \chi_n^{-1}(n_e + n_n)n_\mu \\
    &- \chi_p^{-1}(n_e(n_p - n_\mu) + n_p n_\mu))(\chi^{-1})_e^e + n_e^2 ((\chi^{-1})_e^e)^2] \\
    &\times (\chi^{-1})_\mu^\mu+ n_\mu^2 (\chi_n^{-1} + \chi_p^{-1} + (\chi^{-1})_e^e)((\chi^{-1})_\mu^\mu)^2\Big\}.
\end{split}
\end{equation}
Note that in this manuscript all our numerical results were obtained taking $\lambda_3 = 0$, and thus both the expressions of the relaxations times and partial viscosity coefficients simplify as compared to the general expressions we provide in this Appendix.

The resulting expression in Eq.~\eqref{eq.bulk_viscosity_eff} is consistent with the result reported in Eq.(46) of Ref.~\cite{Alford_2023} with the following identifications

\begin{equation}
\begin{aligned}
    n_1&= \frac{\zeta_+ + \zeta_-}{(\tau_+ \tau_-)^2},\\
    n_2&= \frac{\zeta_+}{\tau_+^2} + \frac{\zeta_-}{\tau_-^2}, \\
    d_1&= \frac{1}{\tau_+ \tau_-},\\
    d_2&= \left(\frac{\tau_+ +\tau_-}{\tau_+ \tau_-}\right)^2.
\end{aligned}
\end{equation}

We have checked that our expressions reproduce the values of the above parameters given in Eqs. (47)-(50) of Ref.~\cite{Alford_2023} for $\lambda_3 = 0$.

\section{Peak value of the bulk viscosity }
\label{sec-app-peak}

It is not easy to determine analytically peak values of the frequency-dependent bulk viscosity, except in some limiting situations. Taking $\lambda_3=0$ and if we further assume that muon densities are frozen
we find 
\begin{equation}\label{eq.npeBV}
    \lim_{\lambda_2\to 0}\zeta(\omega)=\frac{\lambda_1C^2}{\lambda_1^2(\chi_1^{-1})^2 +\omega^2},
\end{equation}
where $C\equiv n_e(\chi^{-1})_e^e-n_n\chi^{-1}_n-n_p\chi^{-1}_p$, which matches the expression of the bulk viscosity of nuclear matter composed only of neutrons, protons and electrons \cite{Harris:2024evy}
\begin{equation}\label{eq.npeBVwithgamma}
     \lim_{\lambda_2\to 0}\zeta(\omega)=\frac{C^2}{\chi_1^{-1}} \frac{\gamma_e(T)}{\gamma_e^2(T)+\omega^2},
\end{equation}
with $\gamma_e(T)\equiv \chi_1^{-1} \lambda_1(T)$. In the temperature regime considered, it is correct to take the EoS at $T=0$ so that all the temperature dependence is on the rates, and thus on $\gamma_e(T)$. In this case it is easy to find that the peak value of the viscosity occurs at $\omega = \gamma_e$, and then \cite{Harris:2024evy}
\begin{equation}
     \left[\lim_{\lambda_2\to 0}\zeta(\omega)\right]_{\rm max}=\frac{C^2}{2\chi_1^{-1}\gamma_e}.
\end{equation}

Finding the peak values of the bulk viscosity \eqref{eq.bulk_viscosity_eff_new} as a function of $T$ does not  allow us to find such a simple expression, as all the second-order transport coefficients depend on $T$. Numerical values of the peak values are however found in the manuscript.  

\clearpage

\bibliography{bibliography}

\end{document}